\documentclass[reprint]{elsarticle}
\usepackage{amssymb,amsthm,amsmath,float}

\usepackage{multirow} 

\usepackage[total={6.5in,8.75in}, top=1.2in, left=0.9in, includefoot]{geometry}
\usepackage{graphicx}
 \usepackage[T1]{fontenc}    

\usepackage[pdftex,bookmarks, colorlinks, citecolor =blue, breaklinks]{hyperref}
\usepackage{url}
\usepackage{xcolor}

\newcommand{\Eq}[1]{(\ref{eq:#1})}

\newcommand{\Sec}[1]{\S \ref{sec:#1}}
\newcommand{\Fig}[1]{Fig.~\ref{fig:#1}}
\newcommand{\Figs}[2]{Figs.~\ref{fig:#1}-\ref{fig:#2}}

\newcommand{\App}[1]{\ref{app:#1}}

\newcommand{\WB} {\mathit{WB}}

\newcommand{\InsertFig}[4]
{\begin{figure}[h!t]
       \centerline{
         \includegraphics[width=#4\columnwidth]{./figures/#1}
       }
       \caption{{\footnotesize  #2}
       \label{fig:#3}}
\end{figure}}

\newcommand{\InsertFigTwo}[5] {
\begin{figure}[h!t]
       \centerline{
         \includegraphics[width=#5\columnwidth]{./figures/#1}
         \hskip 0.5in
         \includegraphics[width=#5\columnwidth]{./figures/#2}
       }
       \caption{{\footnotesize  #3}
       \label{fig:#4}}
\end{figure}}

\newcommand{\InsertFigThree}[6] {
\begin{figure}[h!t]
       \centerline{
         \includegraphics[width=#6\columnwidth]{./figures/#1}
         \hskip 0.1in
         \includegraphics[width=#6\columnwidth]{./figures/#2}
         \hskip 0.1in
         \includegraphics[width=#6\columnwidth]{./figures/#3}
       }
       \caption{{\footnotesize  #4}
       \label{fig:#5}}
\end{figure}}

\newcommand{\InsertFigFour}[7] {
\begin{figure}[h!t]
       \centerline{
\renewcommand{\arraystretch}{0.01}
         \begin{tabular}{cc}
         \includegraphics[width=#7\columnwidth]{./figures/#1}&  \includegraphics[width=#7\columnwidth]{./figures/#2} \\
        \includegraphics[width=#7\columnwidth]{./figures/#3}  &  \includegraphics[width=#7\columnwidth]{./figures/#4}
        \end{tabular}
       }
       \caption{{\footnotesize  #5}
       \label{fig:#6}}
\end{figure}}

\newcommand{\bN}{{\mathbb{ N}}}
\newcommand{\bQ}{{\mathbb{ Q}}}
\newcommand{\bR}{{\mathbb{ R}}}

\newcommand{\bT}{{\mathbb{ T}}}
\newcommand{\bZ}{{\mathbb{ Z}}}

\newcommand{\cO}{{\cal O}}
\newcommand{\cR}{{\cal R}}

\newcommand{\digT}{\mathop{\mathrm{dig}}\nolimits_T}

\newcommand{\beq}[1]{\begin{equation}\label{eq:#1}}
\newcommand{\eeq}{\end{equation}}

\newenvironment{se}[1]{\equation\label{eq:#1}\aligned}{\endaligned\endequation}
\newcommand{\bsplit}[1]{\begin{se}{#1}}
\newcommand{\esplit}{\end{se}}

\newenvironment{example}[1][]
  {
	\setlength \leftmargini {1.0em}		
	\setlength \topsep {0.5em}			
	\begin{quote}
	{\it Example#1} }
	{\end{quote}
  }
\newcommand{\bexam}[1][:]{\begin{example}[#1]}
\newcommand{\eexam}{\end{example}}

\begin{document}
\maketitle

\begin{frontmatter}

\begin{highlights}
\item We study the dynamics of quasiperiodically forced circle maps.
\item We identify strange nonchaotic attractors as invariant sets that are ``weakly chaotic"  using
the weighted Birkhoff average of rotation vector and Lyapunov exponent as computational tools. 
\item Efficient numerics distinguishes between strongly chaotic, weakly chaotic, incommensurate, and resonant orbits.
\end{highlights}

\title{Resonance and Weak Chaos in Quasiperiodically-Forced Circle Maps}
\author[1]{J.D.~Meiss\fnref{fn1}}
\ead{jdm@colorado.edu}
\author[2]{E. Sander\fnref{fn2}}
\ead{esander@gmu.edu}

\fntext[fn1] {JDM was supported in part by the Simons Foundation under Award 601972.
        Useful conversations with Nathan Duignan are gratefully acknowledged.}
\fntext[fn2] {ES was supported in part by the Simons Foundation under Award 636383.}
\affiliation[1] {
 organization = {Department of Applied Mathematics},
 addressline = {University of Colorado},
 city = {Boulder, CO},
 postcode = {80309-0526},
 country = {USA}
 }
\affiliation[2]{
 organization = {Department of Mathematical Sciences},
 addressline = {George Mason University},	
 city = {Fairfax, VA},
 postcode = {22030},
 country = {USA}
 }

\journal{CNSNS}

\begin{abstract}
In this paper, we focus on a numerical technique, the
weighted Birkhoff average (WBA) to distinguish between
four categories of dynamics for quasiperiodically-forced circle maps.
Regular dynamics can be classified by rotation vectors,
and these can be rapidly computed to machine precision using
the WBA. Regular orbits can be resonant or
incommensurate and we distinguish between these by computing their ``resonance order.''
When the dynamics is chaotic the WBA converges slowly.
Such orbits can be \textit{strongly} chaotic, when they have a positive 
Lyapunov exponent or \textit{weakly} chaotic, when the maximal Lyapunov exponent is zero. 
The latter correspond to the strange nonchaotic attractors (SNA) that
have been observed in quasiperiodically-forced circle maps beginning with 
the models introduced by Ding, Grebogi, and Ott. 
The WBA provides a new technique to find SNAs, and allows us to accurately compute the proportions of each of the four orbit types as a function of map parameters.
\end{abstract}


\end{frontmatter}

\tableofcontents 
\date{\today}

\section{Introduction}

It has been conjectured that for typical dynamical systems three types of  
maximal invariant sets are observable:
periodic, topological tori with quasiperiodic dynamics, and chaotic \cite{Sander15}. 
Here we use weighted Birkhoff averages (WBA) \cite{Das16b,Das17,Sander20},
to study this question for quasiperiodically forced circle maps.
The WBA allows for quick and accurate distinction between regular and chaotic orbits
and also provides a highly accurate calculation of rotation vectors for
regular orbits---it is super-polynomially convergent for Diophantine vectors \cite{Das18b}.
Given an accurate computation of a frequency, the method of resonance orders \cite{Meiss21} 
distinguishes between those that are numerically incommensurate and commensurate.
Combining these techniques, we are able to quickly classify orbits as chaotic,
resonant, or quasiperiodic on a torus in one or more dimensions. 

Alternative algorithms for computing rotation vectors and sets include \cite{Alseda21} for circle maps, the 
set-based methods in \cite{Polotzek17} for torus maps, and methods for
numerical continuation of invariant tori \cite{Sanchez10}.
The parameterization method can be used to explicitly compute the conjugacy to rigid rotation \cite{Haro16},
and the needed rotation vector can be efficiently computed using the WBA \cite{Blessing23}.
The frequency analysis method of Laskar \cite{Laskar93a,Laskar03} uses a Hanning window to give
a quadratically convergent Fourier amplitudes to compute rotation vectors.
In a series of papers Villanueva and collaborators use Richardson extrapolation to estimate
rotation numbers for analytic circle diffeomorphisms \cite{Seara06, Luque08, Seara09, Luque14, Villanueva22}.
Experimentally, this appears to give
super-convergence for Diophantine irrationals, but the convergence has not been rigorously shown.
For a comparison of many of these methods to the WBA, see the discussion in~\cite{Das17}.

In this paper, we use the WBA to distinguish between
regular and chaotic orbits for quasi\-periodically-forced circle maps.
Though these are two-dimensional they still retain
some of the behavior of one-dimensional maps; for example, when the map
is a homeomorphism then every orbit has the same rotation vector
\cite{Herman83b}. These maps can have strange nonchaotic
attractors (SNA), i.e., geometrically ``strange'' attractors that have
nonpositive Lyapunov exponents \cite{Ding89, Sturman99, Osinga01, Glendinning09}.
In the classic examples \cite{Herman83b, Grebogi84}, SNAs are weakly chaotic---exhibiting sensitive
dependence on initial conditions but not exponential divergence \cite{Glendinning06}.
It was previously shown that the WBA identifies SNAs for the
forced-damped pendulum \cite{Duignan23}. We show here that 
the WBA identifies strange nonchaotic attractors in forced circle maps.


The main focus of the paper is to use the WBA to distinguish between regular
(either incommensurate or resonant) and chaotic behavior.
In \Sec{Theory} we briefly recall some of the theoretical background for torus maps.
The appendices recall the numerical methods we developed in \cite{Sander20,Meiss21}: \App{wba} explains how to use
the WBA to compute rotation vectors and how it gives an efficient method to identify
chaos. In \App{ResonanceOrder} we recall the method of ``resonance orders''
to distinguish resonant (lower-dimensional) invariant tori from those that are nonresonant
(full-dimensional). 
We apply these techniques in \Sec{QPForcing} to classify the orbits of two-dimensional
torus maps with a rigidly rotating second component.
In \Sec{SNA} we show how the WBA together with Lyapunov exponents
can be used to distinguish strongly chaotic orbits from those that are weakly chaotic.
In \Sec{statistics} we compute the proportions of the four categories of orbits---resonant,
incommensurate, weakly chaotic and strongly chaotic---as a function of the parameters of the map.
We conclude in \Sec{Conclusions}. 

\section{Torus Maps: Background}\label{sec:Theory}

We begin by recalling some of theoretical background for
torus maps $f: \bT^d \to \bT^d$ that are homotopic to the identity.
If $\pi: \bR^d \to \bT^d$ is the standard projection, 
then a map  $F: \bR^d \to \bR^d$ is a lift  of $f$ if
\[
	\pi \circ F  = f  \circ \pi .
\]
Here we take the period of the torus to be one, so that $F(x) \mod 1 = f( x \mod 1)$.
Since $f$ is homotopic to the identity, $F(x+m) = F(x)+m$ for each $m \in \bZ^d$, i.e.,
the map has degree one. Note that any two lifts, say $F_1$ and
$F_2$, differ by an integer vector $F_1(x) = F_2(x) + m$---indeed this
must be true for any fixed $x$, but by continuity the same vector $m$
must work for all $x$.

In general we can assume that a degree-one torus map has the form
\beq{torusMap}
	x' = f(x) = x +\Omega + g(x;a)  \mod 1 ,
\eeq
where  $\Omega \in \bT^d$, $a$ is a parameter vector, and $g$ is periodic,
$g(x+m;a) = g(x;a)$ for any $m \in \bZ^d$ (and every $a$). 
This simplest case is  Arnold's circle map, where $d=1$, and 
\[
	g(x;a) = \frac{a}{2\pi} \sin(2\pi x).
\] 
In \Sec{QPForcing} we consider quasiperiodically-forced circle maps, 
where $d=2$ and \Eq{torusMap} is of the form 
\[
	f(x_1,x_2) = (f_1(x_1,x_2), x_2 + \Omega_2) \, ,
\]
 with $\Omega_2 \in \bR \setminus \bQ$, i.e., it is a skew product and the second component is a rigid rotation with irrational rotation number.
A commonly studied example extends the Arnold map, using the form \Eq{torusMap} with
\beq{QPForce}
	g(x_1,x_2)  = \tfrac{1}{2\pi} \begin{pmatrix}  a_1 \sin(2 \pi x_1 )+ a_2 \sin(2 \pi x_2 ) \\
								  0 \end{pmatrix} . 
\eeq
Such maps have been studied in \cite{Ding89, Feudel95, Osinga01, Stark02, Kim03,Jager06b, Glendinning09}. 

The  orbit of $x \in \bT^d$  has (pointwise) rotation vector $\omega \in \bT^d$ if the limit
\beq{RotationVector}
	\omega(x, f) = \lim_{t \to \infty} \frac{F^t(x) -x}{t}  \mod 1
\eeq
exists. This is independent of the choice of lift; however,
it can depend upon the initial point.


The quasiperiodically-forced map \Eq{torusMap} with \Eq{QPForce} is a homeomorphism
whenever $|a_1| \le 1$. Herman showed that whenever a skew product map on $\bT^2$ is a homeomorphism,
then every orbit has a rotation number that depends continuously on $f$ and moreover that
the limit \Eq{RotationVector} is independent of $x$ \cite{Herman83b,Stark02}.
When the map is a homeomorphism, the graph $\{ \omega_1(x,f): \Omega_1 \in \bR\}$ for 
fixed  $a_1,a_2,$ and $\Omega_2$, is monotone increasing \cite{Bjerklov09}.

Note that since the quasiperiodic map has a trivial second component, 
\beq{Omega2}
	\omega_2(x,f) =\Omega_2.
\eeq
Moreover, when $a_1 = 0$,
\beq{OmegaAZero}
	\omega_1 = \lim_{t\to\infty} \frac{1}{t}\left( t \Omega_1 + \frac{a_2}{2\pi}\sum_{j=0}^{t-1} 
			    \sin(2\pi (j\Omega_2 + x_2(0))) \right) 
			  = \Omega_1,
\eeq
since the trigonometric terms average to zero. Thus for $a_1 = 0$,
 $\omega(x,f) = \Omega$ for any $a_2$ and any $x$.

More generally, the convergence of the limit \Eq{RotationVector}, if it exists,
can be slow, especially when the orbit is chaotic or 
the system has a strange nonchaotic attractor \cite{Ding89, Feudel95, Stark02}.
Instead of \Eq{RotationVector}, we compute the weighted Birkhoff average
\beq{omegaWBA}
	\omega_T = \WB_T(F(x)-x) = \Omega +  \WB_T(g(x;a)),
\eeq
using $T$ iterates, see \Eq{WB}.
As noted in \App{wba}, the WBA accelerates the convergence of the pointwise rotation vector
\Eq{RotationVector} especially when the orbits are not chaotic.

\section{Regular and Chaotic Orbits}\label{sec:QPForcing}

In this section we study the map \Eq{torusMap} with the force \Eq{QPForce}.
In this paper we think of $\Omega_2$ primarily as a ``structural parameter,'' fixing it to be 
\beq{GoldenMean}
	\Omega_2 = \gamma \equiv \tfrac12 (\sqrt{5} -1) \approx 0.618034,
\eeq
the inverse of the golden mean. It is computationally infeasible to perform a 
detailed parameter study to determine  how the dynamical behavior depends 
on the three remaining parameters $(\Omega_1,a_1,a_2)$.
In most cases, we will study the
dependence upon $(\Omega_1,a_1)$ for fixed $a_2$, but in 
several cases we will instead fix $a_1$ and and vary $(\Omega_1,a_2)$. 

\subsection{Regular Orbits}

Since the $x_2$ dynamics for \Eq{QPForce} is a rigid rotation
and $\omega_2 = \Omega_2 = \gamma$ is irrational, each orbit 
$\{(x_1(t),x_2(t)): t \in \bN \}$ is dense on $x_2 \in [0,1]$.
One way to visualize these dynamics is to take a Poincar\'e section, say at $x_2 = 0$;
however, since the orbit is discrete we instead use a ``Poincar\'e slice,''
plotting points at a sequence of times $t_j$ for which $|x_2(t_j)| < 0.0005$, see \Fig{qpTraj}.
In this figure we use the criterion \Eq{2DCriterion} to select only the \emph{nonchaotic} orbits,
those for which the WBA has converged to at least nine digit accuracy,  $\digT \ge 9$, after $T= 10^6$ iterates.
The vertical axis in the figure is the image
\[
	x_1(t_j+1)  = f_1(x_1(t_j),x_2(t_j)) \approx f_1(x_1(t_j),0), 
\]
which is the Arnold function. Thus when $a_2$ is sufficiently small the resulting figure can be
viewed as a perturbation of the one-dimensional Arnold map.
Indeed, even when $a_2 = 0.6$, in panel (a), the resulting orbits resemble those that one would see
in Arnold's map, and there are many regions with incommensurate frequency vectors (small dots),
separated by resonant tongues, regions where $ m \cdot \omega = n$
for some nonzero $(m,n) \in \bZ^3$ (large dots) see \Eq{ResonantPlanes}. 
Note that a resonant tongue corresponds to the existence of an attracting invariant circle on $\bT^2$.
When $m_1$ and $m_2 \neq 0$ these resonances have irrational $\omega_1$:
\beq{QPResonance}
	\omega_1 = \frac{n}{m_1} - \frac{m_2}{m_1} \gamma \in \bR \setminus \bQ .
\eeq
Computing the resonance order and using criterion \Eq{Incommensurate} for \Fig{qpTraj}(a),
where $a_1 = 0.8$, we find $\omega$ is incommensurate for $26.7\%$ of the orbits,
while it resonant for $72.3\%$  (the remaining $1\%$ are chaotic and not shown here).
By contrast, in panel \Fig{qpTraj}(b) where $a_2 = 2.49$, only $2\%$ are incommensurate 
while $94.7\%$ of the orbits are resonant ($3.3\%$ are chaotic).
Finally for $a_2 = 5$ in panel (c), $1.1\%$ are incommensurate, while $96.8\%$ are resonant ($2.1\%$ are chaotic). 

\InsertFigThree{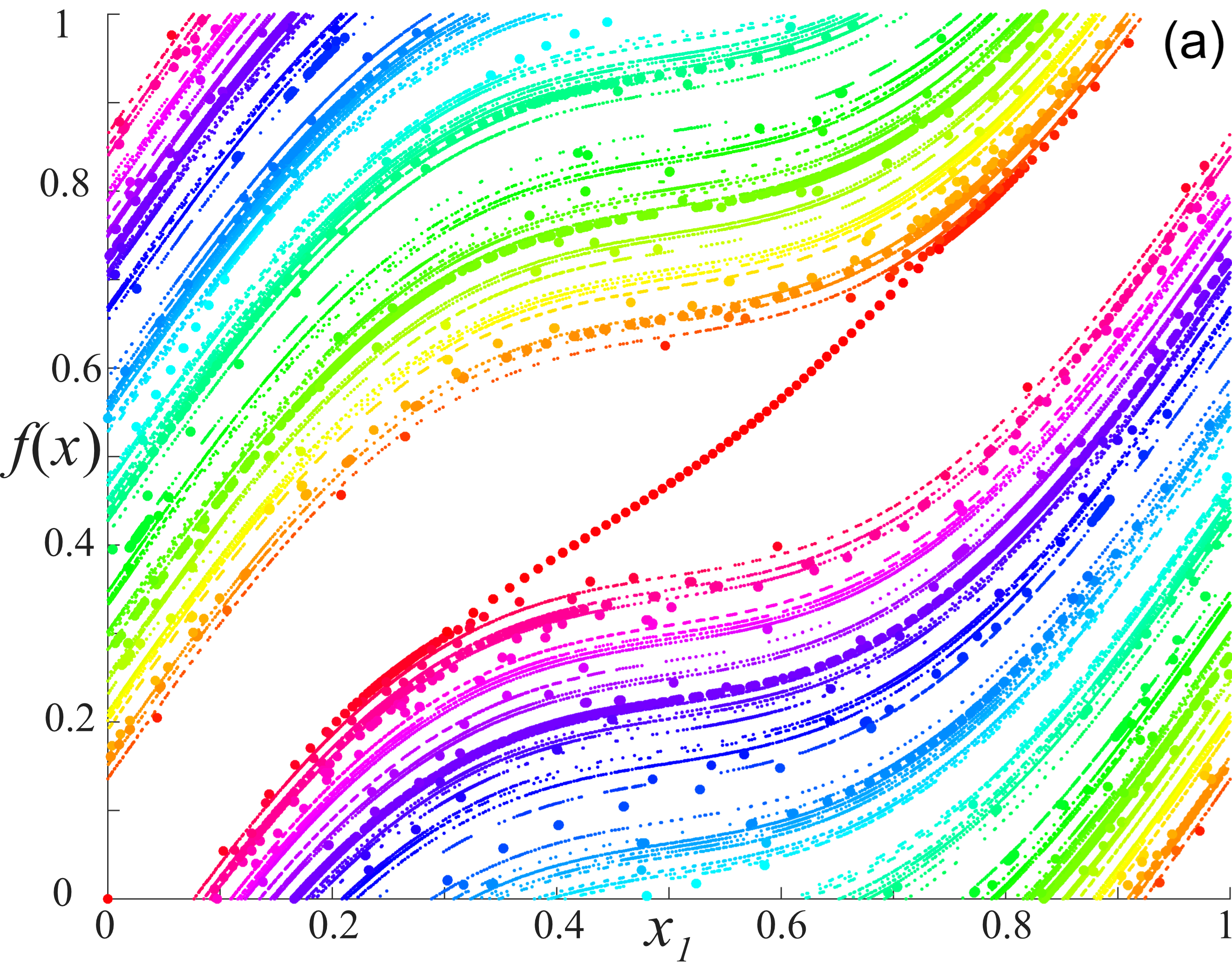}{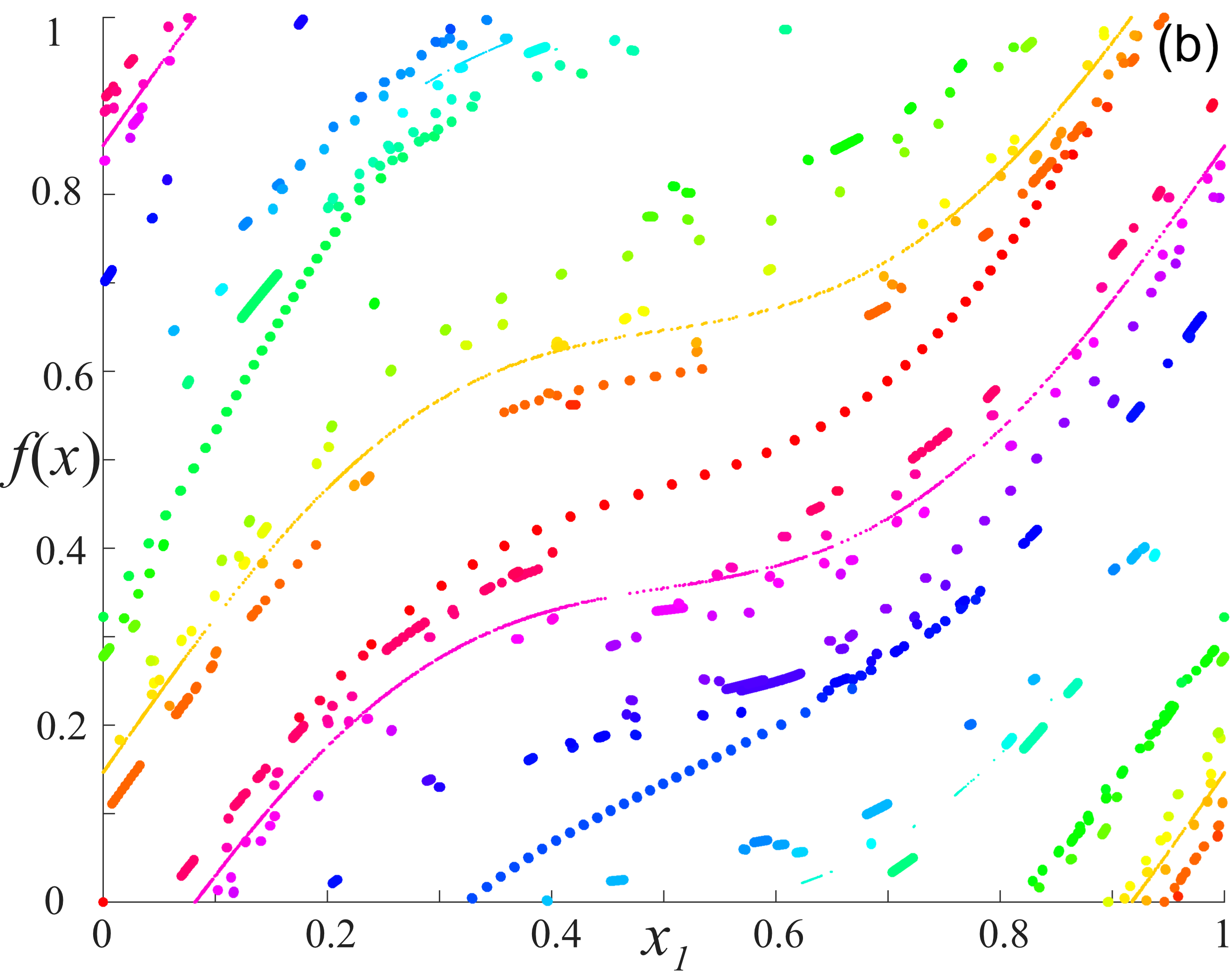}{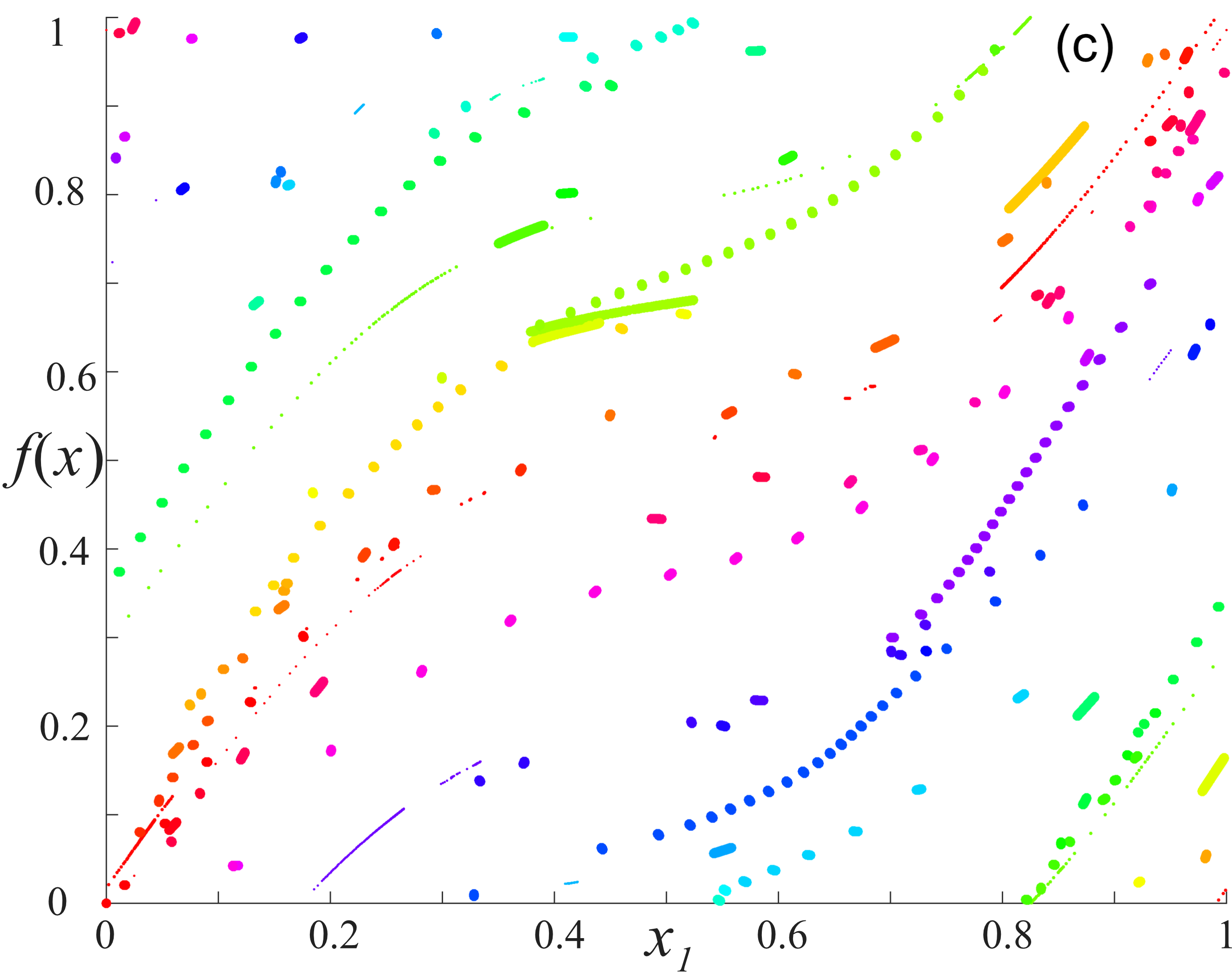} {
Poincar{\'e} slices of regular orbits of the quasiperiodically-forced circle
map with $a_1=0.8$ and $\Omega_2 =\gamma$, for (a) 
$a_2 = 0.6$, (b)  $a_2 = 2.49$ and (c) $a_2 = 5$
(these are the first, fourth, and eighth values shown in \Fig{qpPropa2}). 
The plot shows a grid of 200 for $\Omega_1 \in [0,1]$, 
with the orbits colored as in \Fig{qpMapa1}
using the value of $\omega_1$ computed with $T = 10^6$. Resonant orbits are plotted with larger dots.
Each orbit is iterated $10^5$ times  to remove transients, and the next 1000 points on
the Poincar{\'e} slice $|x_2| < 0.0005$ are shown.
}{qpTraj}{0.33}

The rotation number $\omega_1$ for nonresonant and resonant orbits 
is shown as a function of $(\Omega_1,a_1)$ in \Fig{qpMapa1}
for $a_2 = 0.6$ (top panels) and $a_2 = 1.0$ (bottom panels). 
These use a grid of $1000\times 1000$ evenly spaced values of $(\Omega_1,a_1) \in [0,1]\times[0,2]$,
with $\Omega_1$ slightly shifted away from rationals to avoid resonances at $a_1 = 0$. 
Each orbit begins at the same randomly selected point $(x_1(0),x_2(0))$ and is initially iterated $10^4$
times to remove transients. To compute both $\omega_T$ and $\digT$ we use $T=10^6$.
As noted by \cite{Ding89}, the nonresonant regions (left panels) as a function of
$(\Omega_1,a_1)$, look similar to the circle map case when $a_2$ is small; however,
note that the proportion of the nonresonant orbits appears 
to fall essentially to zero for a value of $a_1$ smaller than $1$, unlike the Arnold case
(see \Sec{statistics}, below).

\InsertFigFour{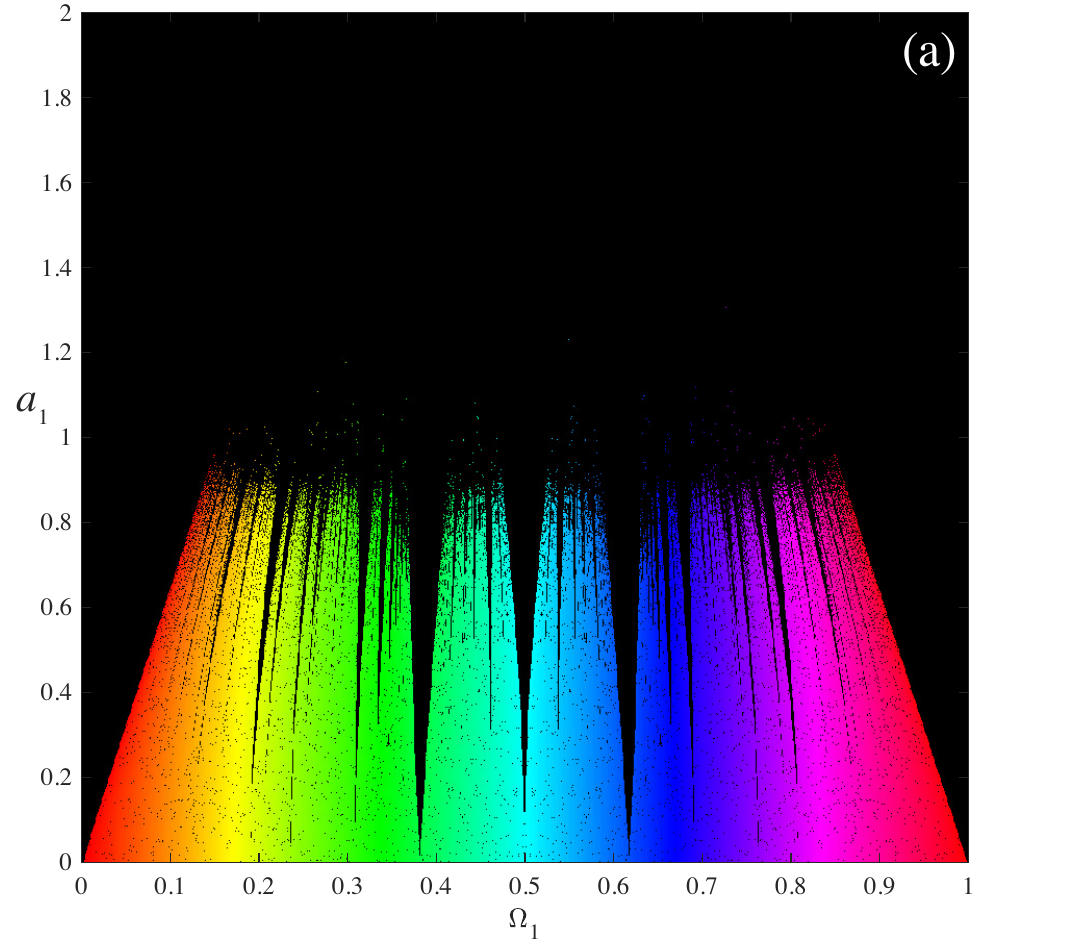}{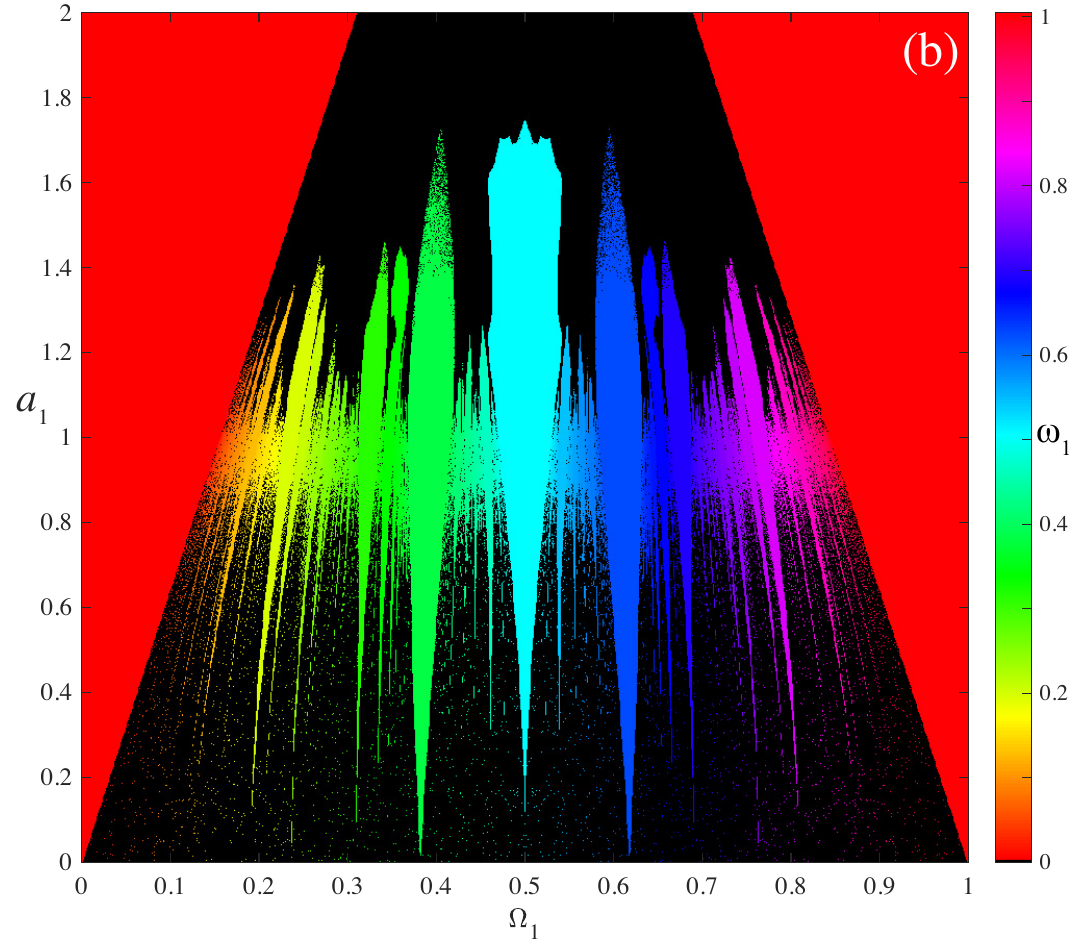}{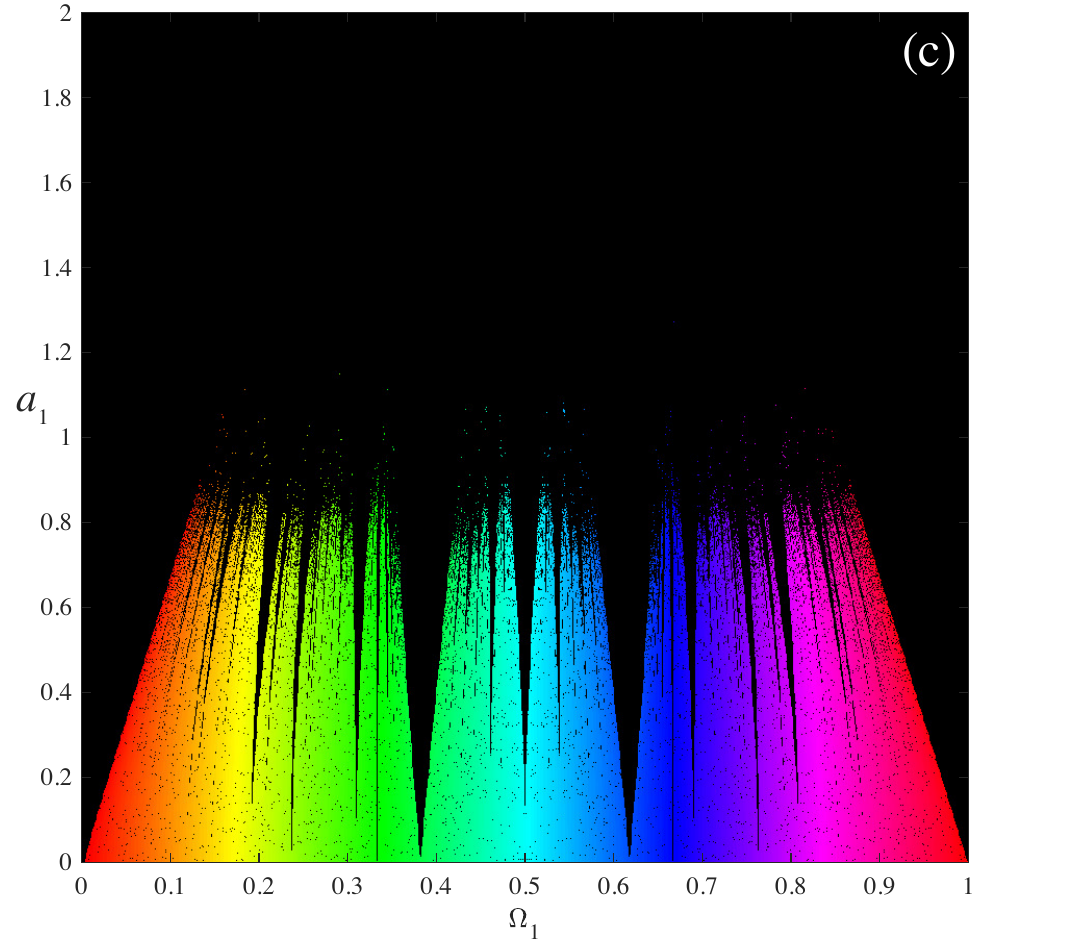}{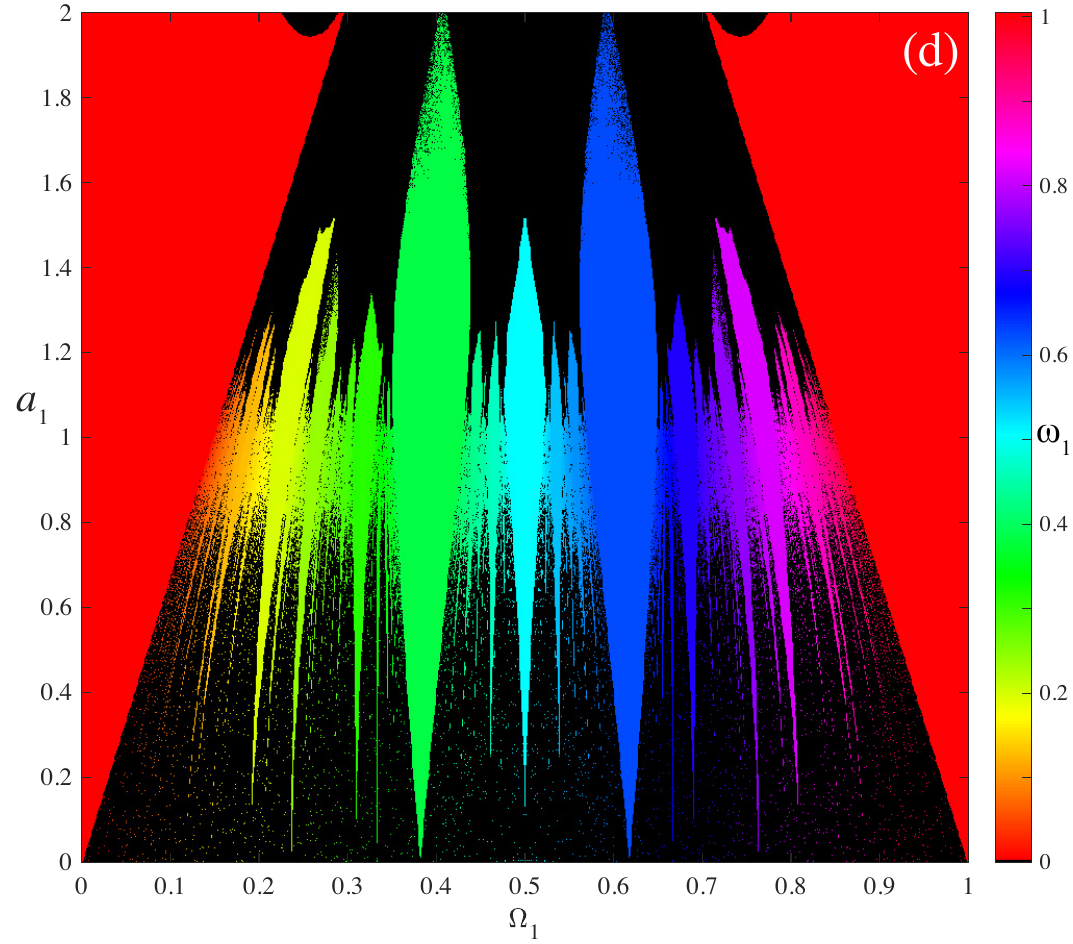}{
Nonresonant (panels (a) and (c)) and resonant (panels (b) and (c)) orbits for
the quasiperiodically-forced circle map \Eq{QPForce}
as a function of $(a_1,\Omega_1)$  for $a_2 = 0.6$ (top panels) and $a_2 = 1$ (bottom panels),
with $\Omega_2 = \gamma$ \Eq{GoldenMean}.
These orbits are distinguished using \Eq{2DCriterion} and \Eq{Incommensurate}.
The orbits are colored using $\omega_1$ as shown in the color bars with
black indicating no orbits of the given type.}
{qpMapa1}{0.5}

\InsertFig{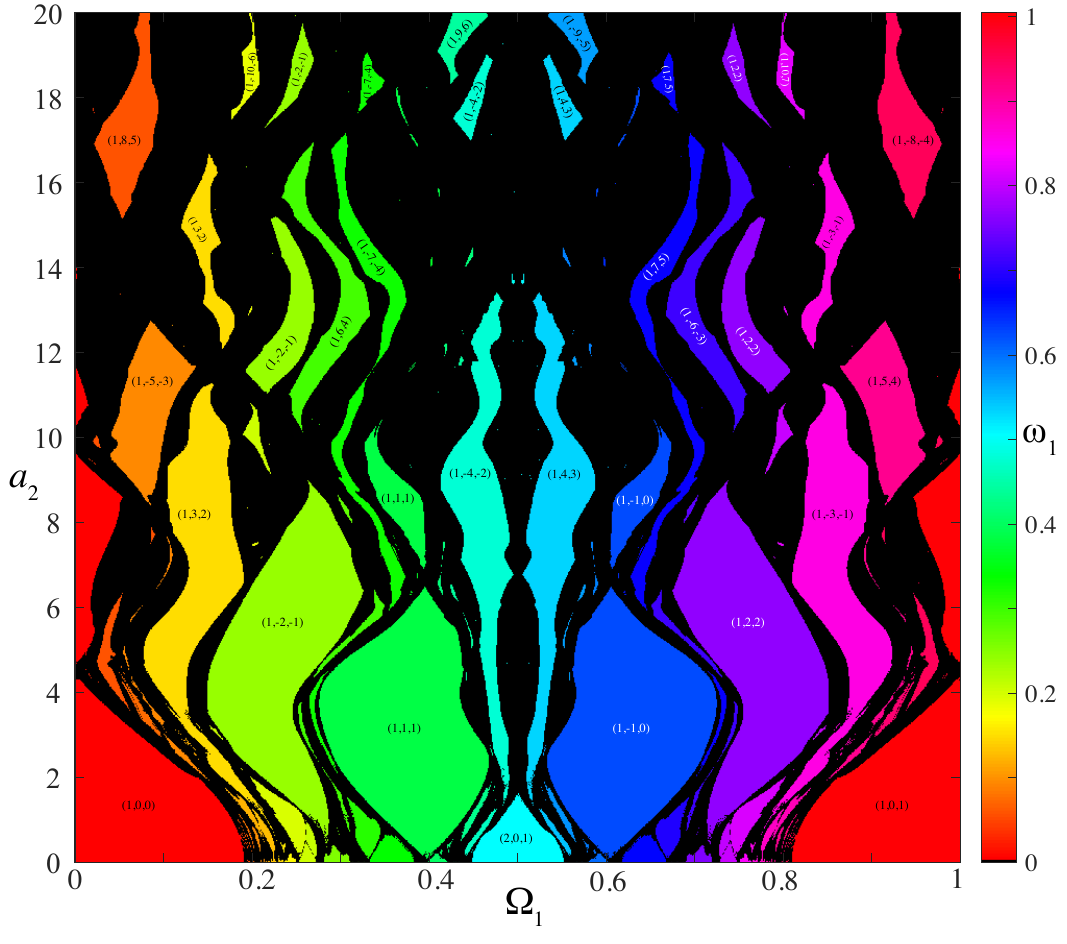}{
Resonant regions for the quasiperiodically-forced circle map \Eq{QPForce} for $a_1  = 1.2$.
The larger regions are labeled by $(m_1,m_2,n)$, Each resonance is colored by $\omega_1$ and black regions correspond
to either chaotic or incommensurate orbits.} 
{qpResonant}{0.75}

The parameters with resonant orbits are shown in \Fig{qpMapa1}(b) and (d) for the same values of $a_2$.
These are analogous to the tongues or mode-locking regions of the Arnold map;
for example, the $\omega_1 = \tfrac{0}{1}$ (red),
$\tfrac{1}{1}$ (red) and $\tfrac{1}{2}$ (cyan) tongues are prominent. 
However, rank-one resonances with $m_2 \neq 0$ \Eq{QPResonance} create additional tongues.
Perhaps the most prominent new mode-locking regions are those with $(m,n) = (1,1,1)$, so that
$\omega_1 = 1- \gamma \approx 0.382$ (green) and with $(m,n) = (1,-1,0)$ so that $\omega_1 = \gamma \approx 0.618$ (blue).
Like the Arnold tongues, these have cusps when $a_1 = 0$ at $\Omega_1 = \omega_1$ and broaden as $a_1$ increases.
Note however that unlike the Arnold map, the resonant regions do not monotonically increase in width when $a_1$ exceeds $1$. 
These ``leaf shaped'' tongues were observed in \cite{Feudel97}. 

An alternative view of the tongues is seen in \Fig{qpResonant}, which shows the resonant regions
for fixed $a_1 = 1.2$ on a $2000 \times 2000$ grid of $(\Omega_1,a_2) \in [0,1]\times [0,20]$.
The larger tongues in the figure are labeled by the computed resonance vector $(m_1,m_2,n)$.
Note that we can accurately compute these regions since, by \Eq{2DCriterion},
we find $\omega_1$ with precision at least $10^{-9}$.
Feudel et al \cite{Feudel97} observed that for $a_1$ fixed, the widths
of the tongues oscillate and narrow over intervals as $a_2$ increases; this is referred to as {\em pinching}.
In \Fig{qpResonant}, some of the resonance tongues appear to pinch off completely, forming a sequence of ``bubbles'' or ``isolas''. However, in general pinching does not require the width to become zero. 
As noted in \cite{Osinga01}, there can be multiple attracting orbits within a tongue (and these can give rise to 
SNAs, see \Sec{SNA}).

In a previous study of resonant tongues, \cite{Osinga01} computed the tongues using bifurcation curves,
concentrating on the case $\omega_1 = 0$, i.e., $(m,n) = (1,0,0)$.
For example, in \cite[Fig. 2]{Osinga01} Osinga et al observe that when $a_1 = 0.8$
the first pinching occurs for $4.1< a_2 < 4.8$;
this corresponds to a region of bistability for several invariant circles.
They note that the formation of the pinches is associated with bifurcations in which a pair of invariant circles collide
at a dense set of points but the collision is not smooth (this is also related to SNAs, see \Sec{SNA}).
In addition, \cite{Feudel97} study the case $a_1 = 1$ with $a_2 \in [0,2\pi]$. Figure 5 of
\cite{Osinga01} shows the $(m,n) = (1,0,0)$ 
tongue first pinches at $a_2 = 4.7$, the $(2,0,1)$ tongue at $a_2 = 3.0$, and $(3,0,1)$ at $a_2 = 0.9$.
Glendinning et al \cite{Glendinning00} show that the width of the $(1,0,0)$ resonance region
is asymptotically
$
	|J_0(a_2/(2\sin(\pi \gamma)))|,
$
for $a_2 \gg 1$ and $a_1 \ll 1$. 
Here $J_0$ is the Bessel function; thus its zeros determine the pinching points in $a_2$, 
and the width approaches zero as $\cO(a_2^{-1/2})$ for large coupling. 
This formula predicts the first pinch at $a_2 = 4.48$. A similar result for $(m,n) = (1,-k,0)$, 
involves the Bessel function $J_k$ instead of $J_0$, reflecting
the zero width of these tongues at $a_2 = 0$. The actual bifurcation
do not always pinch; they can have two or more folds bounding small regions of multi-stability---\cite{Osinga01}
shows that these correspond to saddle-node/pitchfork bifurcations of multiple circles.

Our computations are consistent with these previous results. For example, for $a_1 = 0.8$ (not shown),
we observe that the $(1,0,0)$ tongue first pinches at $a_2 = 4.51$,
though it does not pinch off completely; for $a_1 = 1.2$, the red region in \Fig{qpResonant},
the first pinch point occurs at $a_2 = 4.40$ and the second at $9.63$.
For the $(2,0,1)$ tongue we observe pinch points at $(a_1,a_2) =  (0.8, 2.04)$, $(0.8, 3.96)$, and $(1.2, 1.64)$.
The $(3,0,1)$ tongue is significantly thinner for both $a_1$ values and is not labelled in
\Fig{qpResonant}; when $a_1 = 1.2$, its first pinch point occurs at $a_2 = 0.66$. 

Note that most of the tongues considered in previous works correspond to the special case of $(m_1,0,1)$ resonances.
However, as we see in \Fig{qpResonant}, these are not the most prominent cases:
they are only present for small $a_2$ and appear to disappear entirely as $a_2$ grows.
We also observe that $m_2 \neq 0$ resonances are most prominent when $a_1 = 0.8$ (not shown).

%

\subsection{Chaos}

Orbits that are identified as  chaotic by \Eq{2DCriterion} are shown in \Figs{qpChaotic}{qpChaotica2}.
Instead of using $\omega_1$ for the color scheme---as we did in the previous figures---the colors
indicate the Lyapunov exponent. For the nonlinearity \Eq{QPForce} this is particularly easy to compute since
the Jacobian of \Eq{torusMap} is upper triangular:
\[
	Df = \begin{pmatrix} 1+ a_1 \cos(2\pi x_{1}) & a_2 \cos(2\pi x_{2}) \\
					    0 & 1 \end{pmatrix} .
\]
This implies that one of the Lyapunov exponents is zero and the other is simply the time average
\beq{Lyapunov}
	\lambda = \lim_{t\to \infty} \frac{1}{t} \sum_{j=0}^{t-1} \ln |1+a_1 \cos(2\pi x_1(j))|\;,
\eeq
if this limit exists. Since \Eq{Lyapunov} is the time average of a scalar function on phase space,
we can also easily compute it using \Eq{WB}:
\beq{WBLyap}
	\lambda_T = \WB_T(\ln |1+ a_1 \cos(2\pi x_1)|) \;.
\eeq
The average \Eq{WBLyap} was previously used in \cite{Das17} to show that both Lyapunov
exponents are zero for nonchaotic, nonresonant orbits of a torus map.

\InsertFigTwo{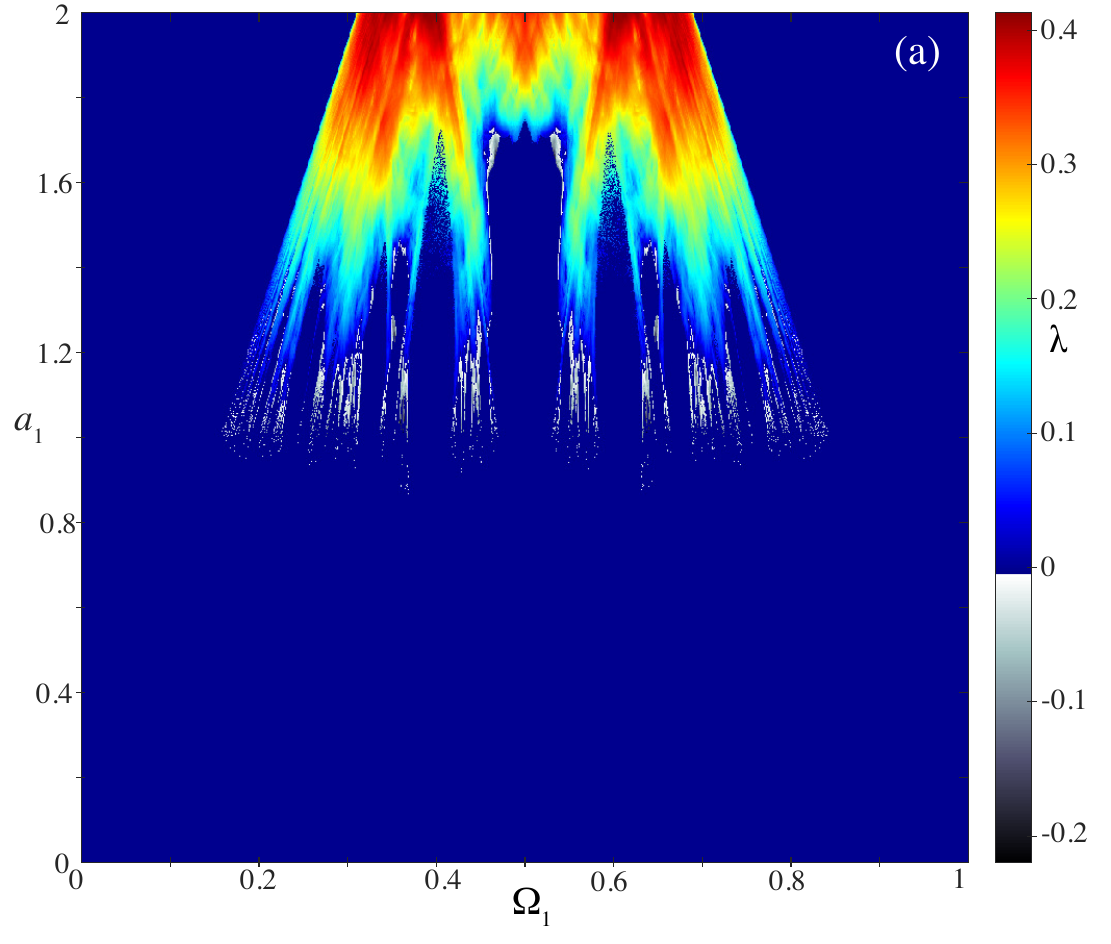}{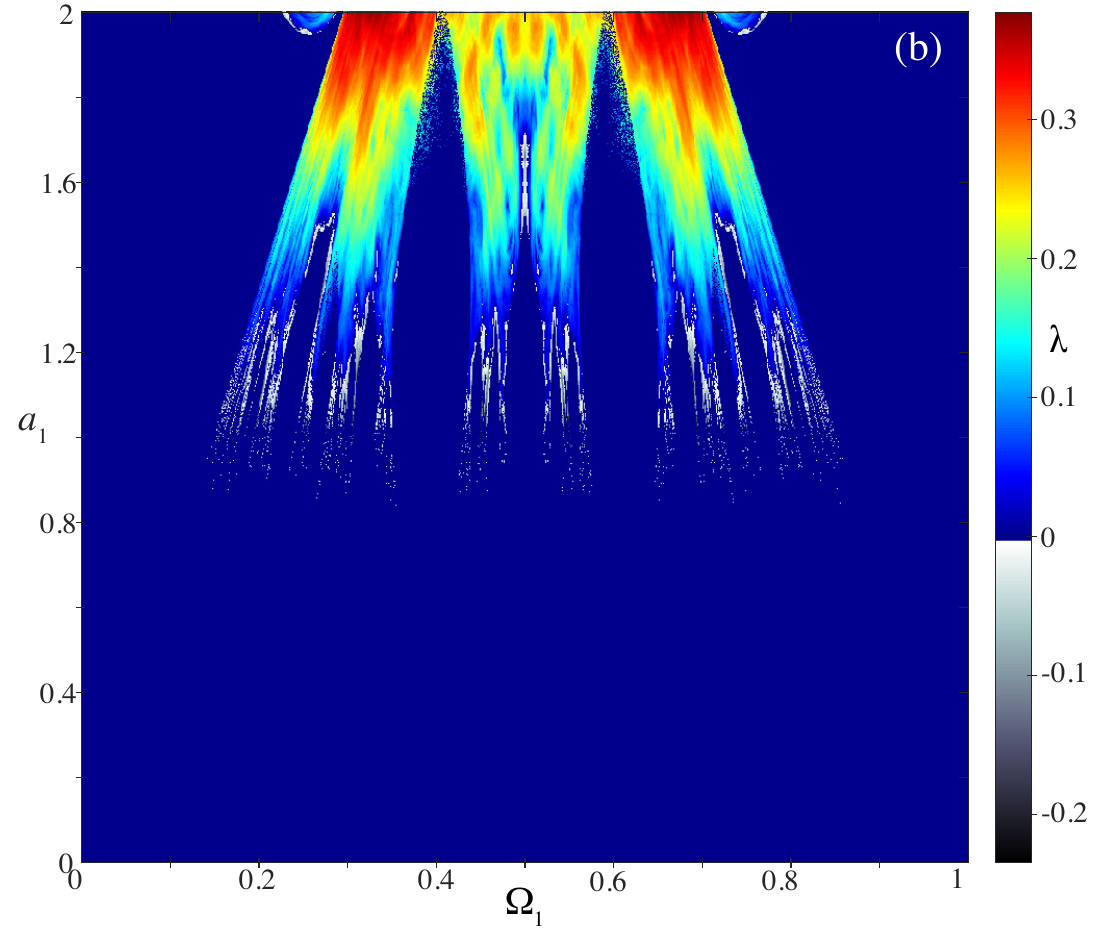}{
Chaotic orbits  of the quasiperiodically-forced circle map \Eq{QPForce} for a $1000 \times 1000$ grid in
the $(\Omega_1,a_1)$ plane for (a) $a_2 = 0.6$ and (b) $a_2 = 1$. The color bar corresponds to $\lambda_T$ \Eq{WBLyap}.
Parameters with $\lambda_T \le 0$ are gray; they correspond to weak chaos, or strange nonchaotic attractors. 
Parameters with nonchaotic orbits are colored dark blue, and strongly chaotic orbits have colors that vary with $\lambda_T$.}
{qpChaotic}{0.5}

Of course, a weighted average like \Eq{WBLyap} should not be expected to improve convergence for orbits that are chaotic; however,
when the orbit is periodic or quasiperiodic, we expect that $\lambda_T$ should converge more rapidly than \Eq{Lyapunov}.
Indeed our tests show that this is the case when $\lambda \le 0$. 
For example, using a $500\times 500$ portion of the grid in 
$(\Omega_1,a_1)$, we find that when $a_2 = 0.6$ approximately $70\%$ have $\lambda_T \le 0$ for $T = 10^6$. 
For these regular cases, if we instead use a much smaller number of iterates, $T= 800$, we found the mean error $\langle \digT\rangle = 3.6$ for \Eq{Lyapunov}, while  $\langle \digT \rangle = 7.9$ for \Eq{WBLyap}.
This more rapid convergence of \Eq{WBLyap} persists as $T$ is increased. 

Note that if $|a_1| > 1$ the function being averaged in \Eq{WBLyap} is not smooth, and
since the weighted average relies on smoothness for improving convergence, we would not expect \Eq{WBLyap} to be helpful.
Moreover, the only case for which the weighted average has been proven to be super-convergent is for orbits conjugate to a rigid quasiperiodic rotation \cite{Das18b}: there are no such orbits when $|a_1| > 1$.
Nevertheless the less accurate computations of the Lyapunov exponent still do indicate
that $\lambda_T > 0$ for many orbits as $a_1$ grows in \Fig{qpChaotic}.
There are also orbits in these panels that are chaotic according to
\Eq{ChaosThreshold}, but for which $\lambda_T \le 0$---they are shown in gray in 
\Figs{qpChaotic}{qpChaotica2}; we discuss these next.

\InsertFig{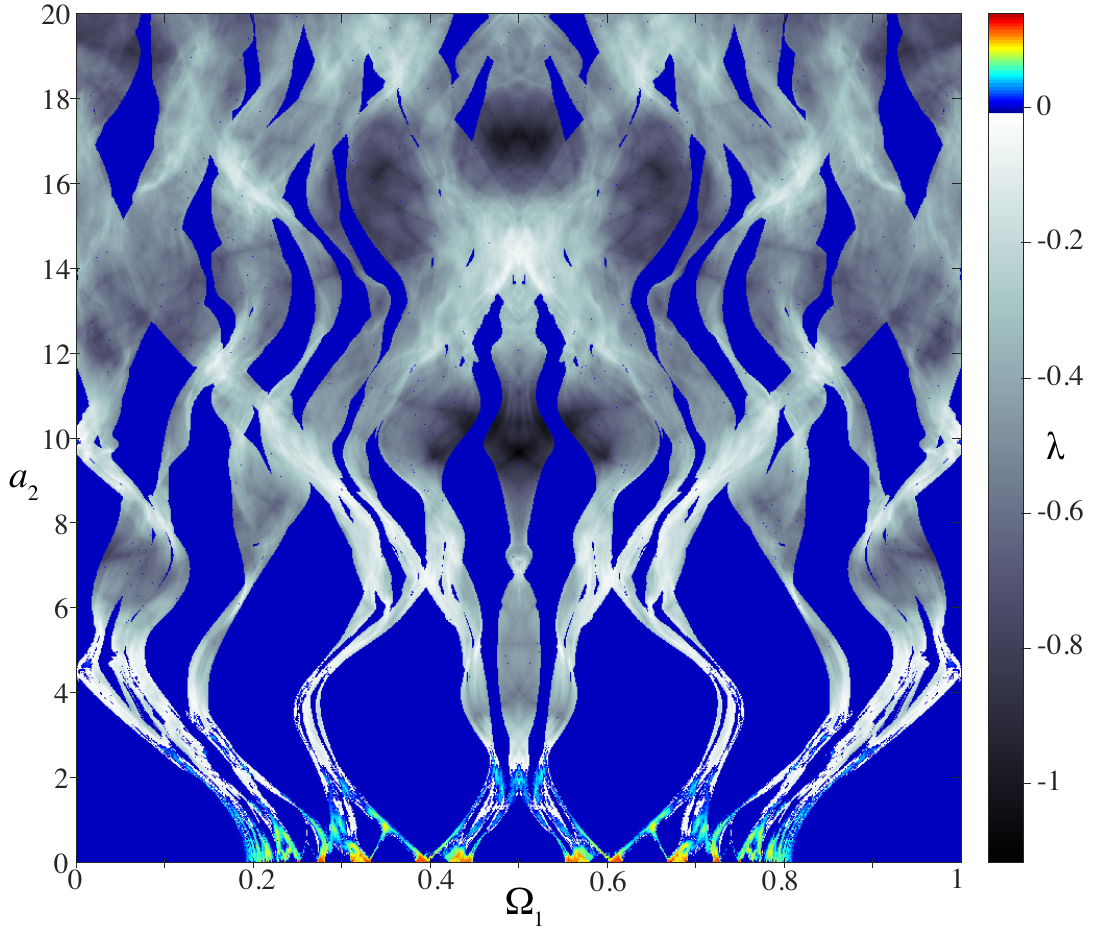}{
Quasiperiodically-forced circle map \Eq{QPForce} for $a_1  = 1.2$ for a $1000\times1000$ grid in $(\Omega_1,a_2)$
showing  orbits with sensitive dependence using Criterion \Eq{ChaosThreshold},
colored by Lyapunov exponent \Eq{WBLyap}.
The grayscale indicates strange nonchaotic attractors. Blue indicates nonchaotic parameters,
most of which correspond to the resonance regions in \Fig{qpResonant}. } 
{qpChaotica2}{0.75}

\section{Weak Chaos}\label{sec:SNA}

It was first observed by \cite{Ding89} that there is a range around $a_1 = 1$ for which
quasiperiodically-forced circle maps can exhibit strange nonchaotic attractors (SNAs),
defined to be attractors that are geometrically strange, but which have $\lambda \le 0$.
In some cases \cite{Glendinning06} showed that
SNAs correspond to ``weak chaos,'' meaning sensitive dependence on initial conditions;
we will use the terms weak chaos and SNA interchangeably.
In particular, the criterion \Eq{ChaosThreshold} combined with negative Lyapunov exponent 
serves as a new criterion for weak chaos/SNA: 
\beq{SNAcriteria}
	\digT<D_T \mbox{ and } \lambda_T \le 0  \Rightarrow \mbox{``weakly chaotic''} \; .
\eeq
Alternatively when $\lambda >0$ we refer to the orbit as having ``strong chaos'', or---more simply---as chaotic.
An example is shown in \Fig{qpSNA}(a); this is an orbit of \Eq{torusMap} and \Eq{QPForce} with
$\digT = 8.4<D_T$ and $\lambda_T = -0.0648<0$, so it satisfies \Eq{SNAcriteria}.
Visually, the phase portrait shows the expected geometric ``strangeness''. 
Since we use the convergence rate of the WBA to detect chaos and not the geometry of the orbit, 
our criterion \Eq{SNAcriteria} differs significantly from previously used methods to identify SNA.


\InsertFigTwo{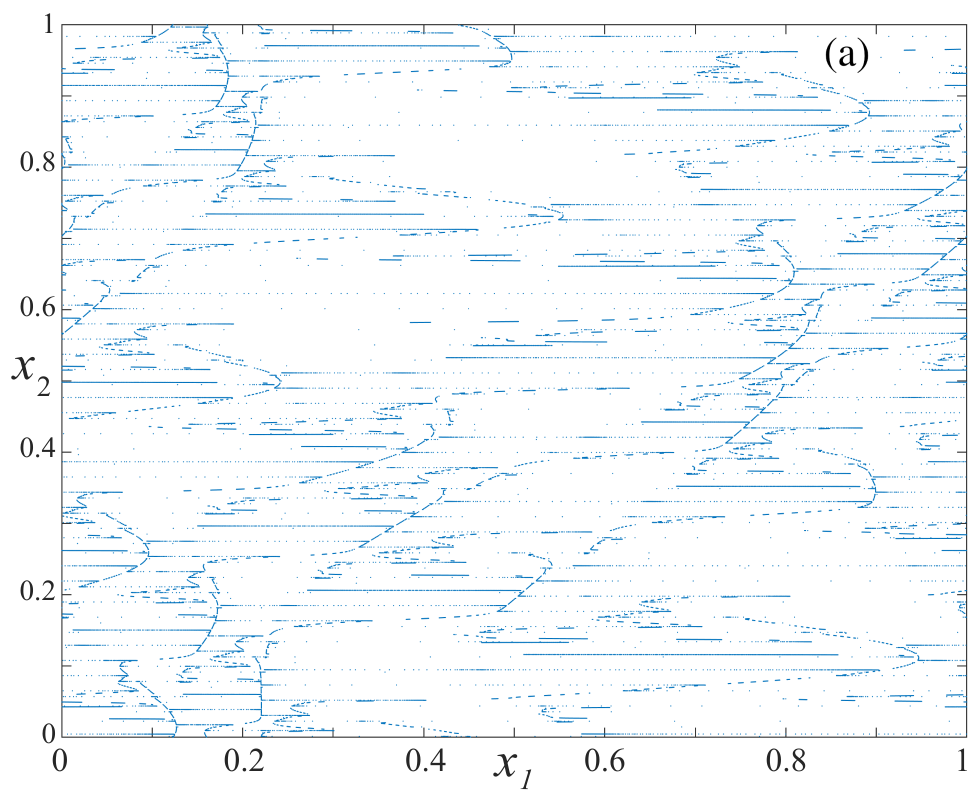}{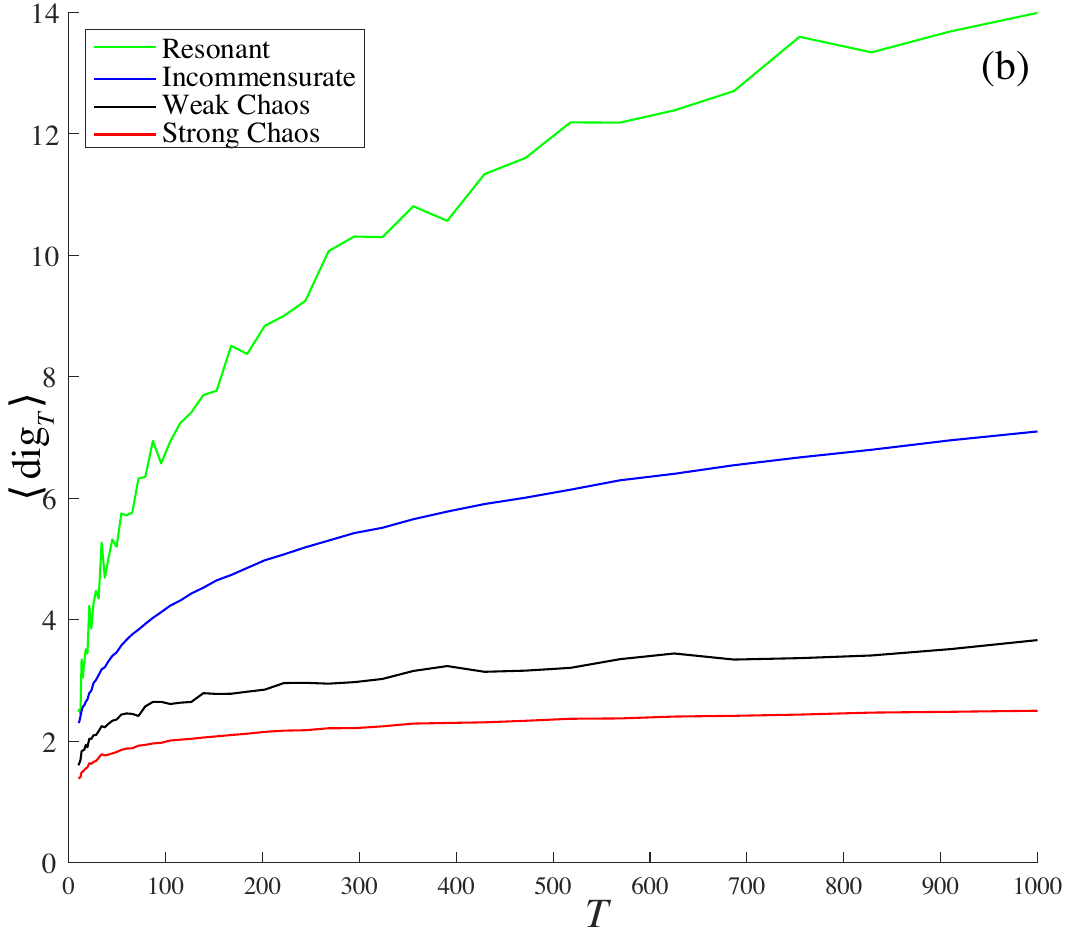} {
(a) A phase portrait of a weakly chaotic orbit for \Eq{QPForce} with $(a_1,a_2) = (0.93493,1)$ and $\Omega_1 = 0.5526$. 
(b) The mean value of $\digT$ as a function of $T \in [0,1000]$ for
\Eq{QPForce} with $a_1 \in [0,2]$, $a_2 = 1$ and $\Omega_1 \in [0,1]$.
The orbits are separated into four types: strongly chaotic orbits
have the smallest $\langle\digT\rangle$ (red), followed by weakly chaotic
orbits (black), nonresonant (blue), and finally resonant orbits (green).
Out of the $10^6$ orbits sampled, $4213$ were identified as weakly chaotic. 
For comparison we randomly selected $4213$ orbits of each of the other types. 
}{qpSNA}{0.5}

For the example in \Fig{qpSNA}(a), $|a_1| < 1$ so the rotation number does exist and is independent of 
initial condition; however, as was emphasized in \cite{Stark02} the convergence of \Eq{RotationVector}
can be slower than $T^{-1}$. Stark et al constructed an algorithm that is guaranteed
to converge as $T^{-1}$ by averaging over initial conditions. To compute this, they use a discrete approximation to the integral on a grid $x_2(0) = \tfrac{i}{N}$, $i = 0,\ldots, N$, to obtain
\beq{OmegaStark}
	\omega^{Stark} = \frac{1}{NT} \sum_{i=0}^{N-1} \left(f^T(x_1,\tfrac{i}{N}) - (x_1,\tfrac{i}{N})\right) .
\eeq
For example, when $(a_1,a_2) = (0.8, 6\pi)$, and $\Omega_1 = 0.01$,
the map has an SNA according to \cite{Feudel95}. Using  $N = 10^3$ and $T = 10^5$, in 
\Eq{OmegaStark}, \cite{Stark02} computed
\[
	\omega_1^{Stark} = 0.0173598 \;.
\]
When we use \Eq{omegaWBA} to compute $\omega_T$ with $T= 10^6$ we obtain the same value to the quoted accuracy, 
and find $\digT =  6.4$,  which is consistent with six digits accuracy. 
For this case,  $\lambda_T = -0.2646098$, so by Criterion \Eq{SNAcriteria} this orbit is indeed weakly chaotic.
This example, and others that we have tried, show that the WBA can give the same 
accuracy as \Eq{OmegaStark} with a factor of $50$ fewer iterates!


We show in \Fig{qpSNA}(b) that there are distinct differences in the number of digits, $\digT$,
for the four categories of orbits:
resonant, incommensurate, weakly chaotic, and strongly chaotic.
Here we selected orbits from the data in \Fig{qpMapa1} and \Fig{qpChaotic}
with $a_2 = 1$.  Of the $10^6$ orbits, $4213$ were identified as weakly chaotic using \Eq{SNAcriteria}.
The black curve in the figure shows the average $\langle \digT \rangle$ for these orbits as a function of $T$.
For a fairer comparison, we randomly selected the same number of orbits of the other three types.
Note that for each $T$ in \Fig{qpSNA}(b), $\langle \digT \rangle$ is ordered monotonically by type;
it is largest for the resonant orbits, smaller for the incommensurate case, smaller for weak chaos and finally smallest for strong chaos.
In particular, $\omega_T$ for the resonant orbits nearly reaches double precision accuracy
(at least in the mean) even for $T = 1000$.
In addition, the rate of increase of $\digT$ with $T$ is also ordered in the same way by type; however,
the distinction in rates between the weakly and strongly chaotic orbits is less pronounced.
Nevertheless, even though these distinctions are clear for the average $\langle \digT \rangle$,
this method does not give a reliable classification: the values of $\digT$ for individual orbits can vary significantly. 

\section{Orbit-Type Statistics}\label{sec:statistics}
The proportions of orbits of each of the four types---strongly chaotic (red), weakly chaotic (black),
resonant (green) and incommensurate (blue)---are shown in \Fig{qpProp}. This figure uses the data
in \Fig{qpMapa1} and \Fig{qpChaotic}, summed over $\Omega_1$, as a function of $a_1$
for two fixed values of $a_2$.

Note that the curves for the incommensurate orbits (blue) should begin at $1$ when $a_1 = 0$ according to \Eq{OmegaAZero};
however, our method underestimates this fraction by about $1.5\%$, declaring that orbits with orders
outside the interval \Eq{Incommensurate} are resonant.
Some cutoff is inevitable, of course, since we cannot compute with infinite precision; moreover, 
a given precision $\delta$ results on average in a smaller resonance order in higher dimensions: compare the factor
of $\tfrac13$ in \Eq{MeanRes} with the factor $\tfrac12$ for the 1D case in \cite{Sander20}.
The incommensurate fraction in  in \Fig{qpProp} drops nearly to zero with 
a shape similar to the power law behavior seen for a circle map \cite{Ecke89, Sander24},
but in contrast to the 1D case, it becomes nearly zero when $a_1$ is below $1.0$;
this corresponds to the onset of weak chaos (black curves).
Of course, when $a_1 >1$, there are no incommensurate orbits, and our method falsely identifies very few.

The fraction of resonant orbits (green) no longer has the cusp at $a_1 = 1$
that is seen for a circle map \cite{Sander24}, and it now decreases more substantially when $a_1 >1$.
This, of course could be anticipated from the right panels of 
\Fig{qpMapa1}, since the widths of some of the resonant tongues decrease for $a_1>1$.

The proportion of weak chaos (black) is largest when $a_1$ is just above $1$: in \Fig{qpProp}(a)
where $a_2 = 0.6$, this fraction is larger when $a_1 \in (1,1.2)$ with a peak of about $10\%$;
the same height is seen in panel (b), where $a_2 = 1$, but the range over which there is weak chaos broadens.
In both cases there is nonzero fraction of weak chaos over a wider range.

The chaotic fraction in \Fig{qpProp} (red) is nonzero for $a_1>1$ and grows to about $40\%$ when $a_1 = 2$,
which is larger than that found for a circle map \cite{Sander24}.

\InsertFigTwo{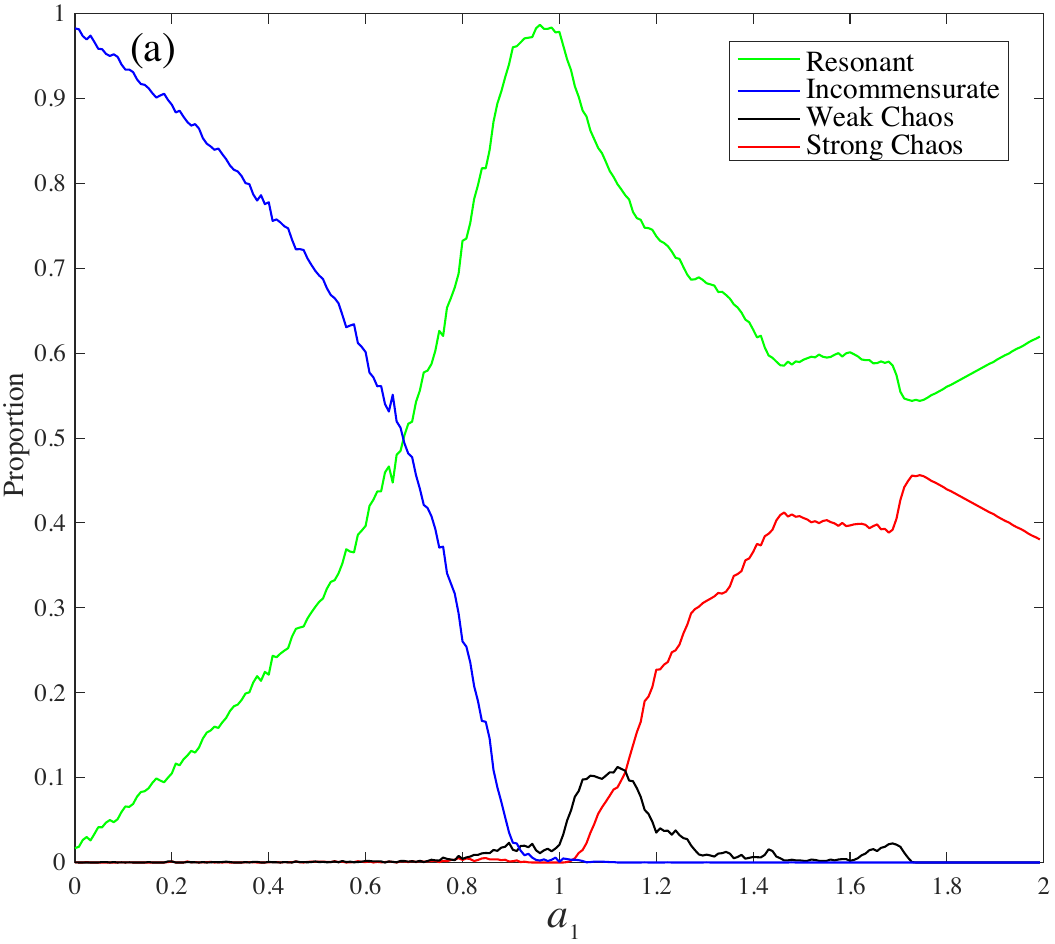}{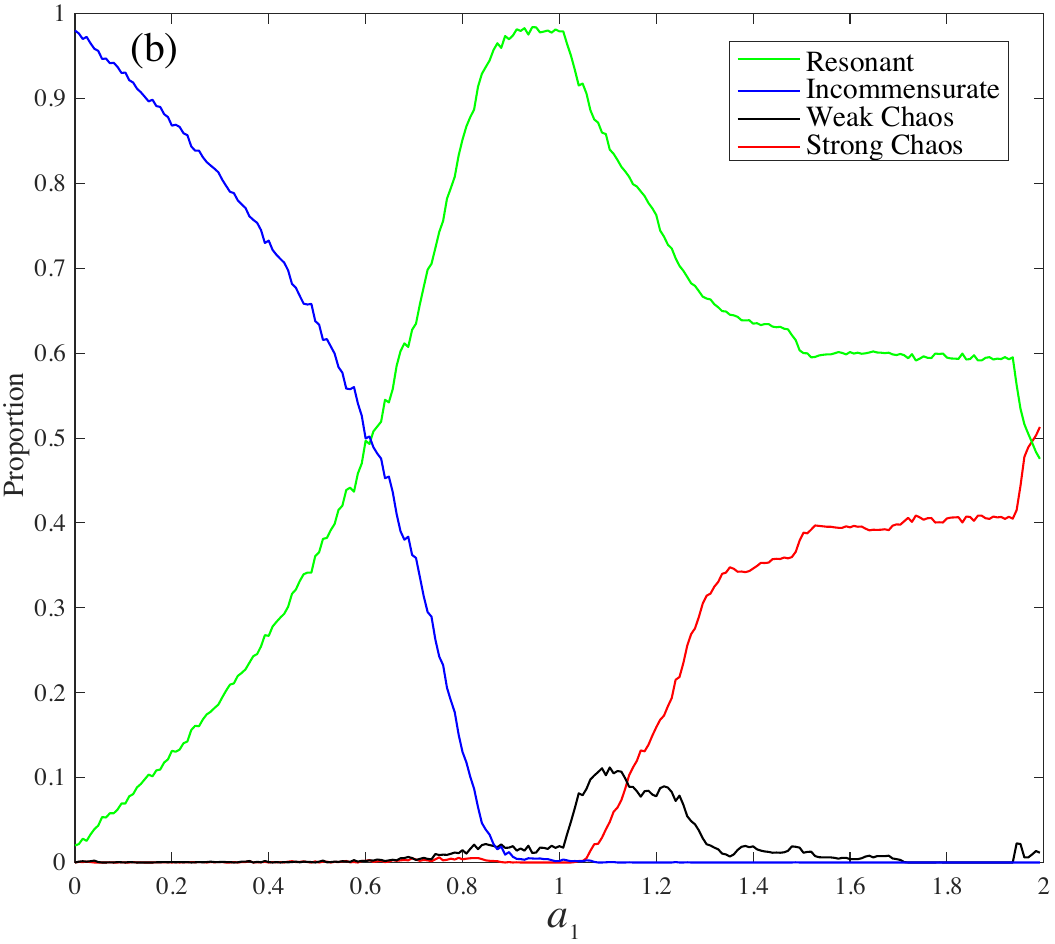}{
The proportion of strongly chaotic (red), weakly chaotic (black), resonant (green), and
incommensurate (blue) orbits for \Eq{torusMap} with \Eq{QPForce} as a function of $a_1$.
(a) $a_2 = 0.6$ and (b)  $a_2 = 1$.  }
{qpProp}{0.5}

Figure~\ref{fig:qpPropa2}(a) shows the parameters exhibiting strong chaos (red)
and weak chaos (black) for eight values of $a_2 \in [0.6,5]$ (vertical axis)
as a function of $(\Omega_1,a_1) \in [0,1] \times [1,2]$ (horizontal axes). 
As was also seen in \Fig{qpChaotica2}, the relative proportion of weak chaos
continues to grow as $a_2$ increases. This three-dimensional view of the data fits
together the viewpoints in  \Fig{qpMapa1} and \Fig{qpChaotic},
which fix $a_2$ and plot $(\Omega_1,a_1)$, and \Fig{qpResonant} and \Fig{qpChaotica2},
which fix $a_1$ and plot $(\Omega_1,a_2)$. 

The proportion of the four orbit types as a function of $a_1$ is shown in \Fig{qpPropa2}(b)
for each of the eight values of $a_2$ from \Fig{qpPropa2}(a).
Here increasing thickness of the curves is used to indicate increasing values of $a_2$. 
As we noted in \Fig{qpProp}, the proportion of incommensurate orbits (blue curves)
are underestimated at $a_1 = 0$ by about $1.5\%$.
Even though the curves in \Fig{qpPropa2}(b) do reach zero at $a_1 = 1$,
it appears that there is no universal power
law analogous to that of circle maps \cite{Ecke89} for these forced maps.
The incommensurate fraction drops to a small value at the onset of weak chaos,
and this onset point decreases as $a_2$ grows.
Moreover the shape of these curves below this onset varies as $a_2$ increases. 
The proportion of weakly chaotic orbits (black curves) grows substantially with $a_2$,
reaching $60\%$ when $(a_1,a_2) \approx (1.75,5)$.
This is at the expense of the resonant orbits: each of 
the green curves peaks near $a_1 = 1$ and decreases thereafter.
The resonant proportion falls to near zero at $(a_1,a_2) \approx (2,5)$ as the chaotic fraction grows.

\InsertFigTwo{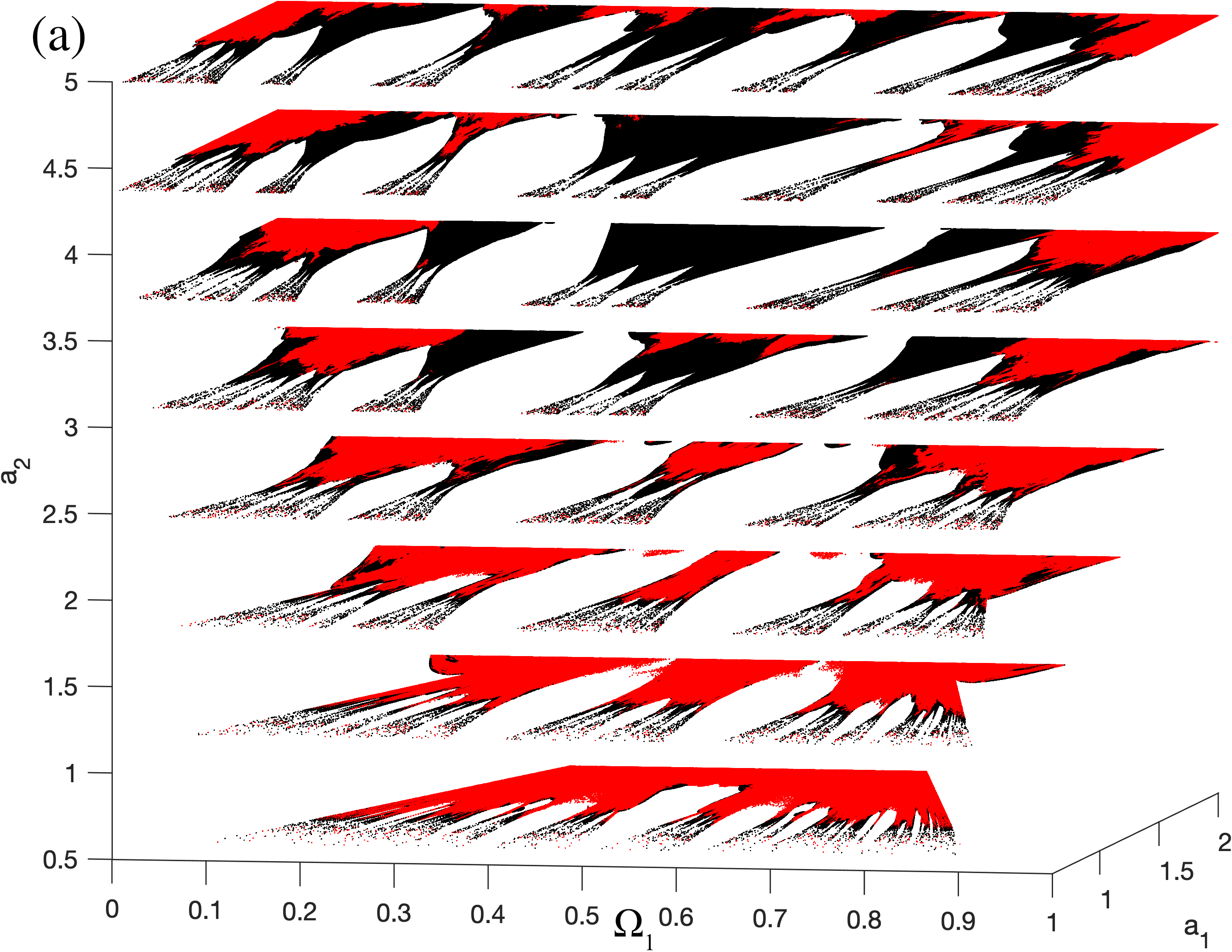}{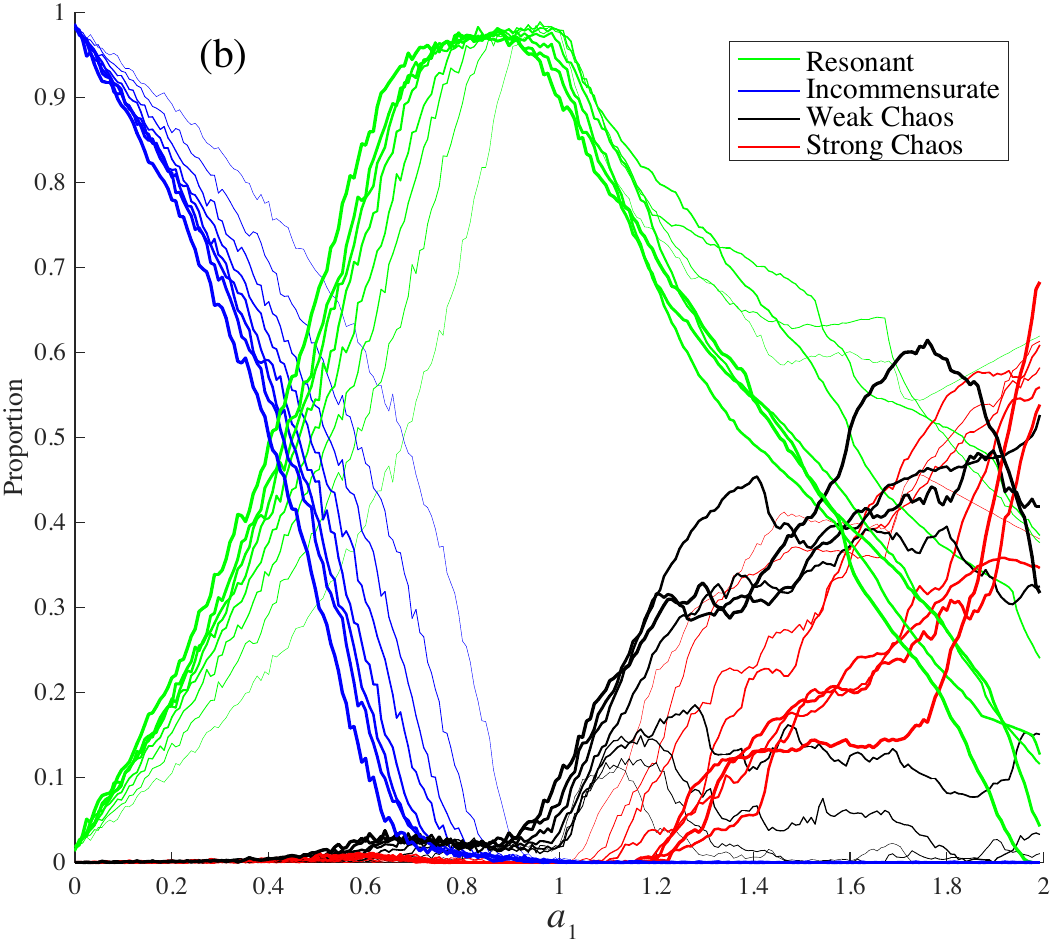} {
Types of orbits for a grid of eight $a_2\in [0.6,5]$ as a function of $(a_1,\Omega_1)$ for 
\Eq{torusMap} with \Eq{QPForce}. 
(a) Parameter regions with strongly (red) and weakly (black) chaotic orbits.
The range $a_1 \in (0.6,1)$ is shown since the chaotic fractions are negligible for smaller $a_1$ values.
(b) Proportion of the four orbit types as a function of $a_1$ for each of the eight $a_2$ values of panel (a)
using the color scheme of \Fig{qpProp}.
The thickness of the eight curves increases with $a_2$. 
}{qpPropa2}{0.5}

\section{Conclusions}\label{sec:Conclusions}

In this paper, we used efficient, high-precision numerical techniques to characterize the dynamics of
a quasi\-periodically-forced extension of the Arnold circle map. An advantage of the weighted Birkhoff
average (WBA) \Eq{WB} is that it can rapidly compute the rotation vector, as well as
the time average of other phase space functions \cite{Sander20,Meiss21}, to near machine precision
when the orbit is regular: it is super-convergent when the orbit is smoothly conjugate
to a rigid Diophantine rotation \cite{Das18b}. We observed that the computations are even more rapidly
convergent when the rotation vector is resonant, recall \Fig{qpSNA}(b).

We used the WBA to distinguish between chaotic and regular
orbits by defining a threshold for the precision, $\digT$ \Eq{digits}, after a fixed number of iterates $T$.
Given an accurate value for rotation vector $\omega_T$ \Eq{omegaWBA}, we determined if the vector is nearly resonant by finding the smallest order of an integer resonance plane within a distance $\delta$. This allows us
to characterize regular orbits as incommensurate (nonresonant) or rank-one (resonant).
We used a similar, but more efficient method---based on the Farey tree---for a scalar frequency in \cite{Sander20}. 
We hope that an efficient algorithm to compute \Eq{Mres} for higher dimensional cases can be found. 
Meanwhile, we used a brute force method, following our previous work on volume-preserving maps \cite{Meiss21}.
The $(m,n)$ resonances correspond to tongues or mode-locking regions, recall \Fig{qpResonant},
and we observed that the pinching of these as a function of the coupling parameter, previously
reported for low-order resonances, also occurs at higher orders as well as for nonzero $m_2$.

Our method gives a new characterization of strange nonchaotic attractors (SNA)
and an efficient way to distinguish these from chaos, recall \Figs{qpChaotic}{qpChaotica2}.
We identify SNAs as invariant sets that are ``weakly chaotic'':
orbits for which the WBA converges slowly but the Lyapunov exponent is not positive.
We showed that the WBA can also be used to improve the accuracy of Lyapunov exponents
over the simplest method \Eq{Lyapunov}---though this does not give
the super-convergence seen for Diophantine rotations. We hope
to investigate the computation of Lyapunov exponents in more detail in a future paper.

Using the method above, we were able to efficiently classify the trajectories of quasi\-periodically forced
circle maps into four categories: resonant,
incommensurate, weakly chaotic, and strongly chaotic, and---as in \Figs{qpProp}{qpPropa2}---showed how
their proportions vary with the strength of the nonlinearity. 

We plan to report on the application of these methods to fully coupled torus maps in another paper \cite{Sander24}.

\newpage
\appendix
\begin{center}
\Large{\textbf{Appendices}}
\end{center}

\section{Weighted Birkhoff Averages}\label{app:wba}
We briefly review here the weighted Birkhoff average \cite{Das16a, Das17, Das19} and how to use it
to distinguish between regular and chaotic orbits \cite{Sander20}. Given a map $f:M \to M$, recall that the time average of a function $h: M \to \bR$ along an orbit of $f$ is simply
\beq{Birkhoff}
	B(h)(z) = \lim_{T \to \infty} \frac{1}{T} \sum_{t=0}^{T-1} h \circ f^t(z) ,
\eeq
if this limit exists.
Under the assumptions that $\mu$ is an $f$-invariant probability measure
($\mu(A) = \mu(f^{-1}(A))$ for any Borel subset $A$ of $M$),
$\mu$ is ergodic, and $h \in L^1(M,\bR)$, then  Birkhoff's ergodic theorem implies that 
\[
	B(h)(z) = \langle h \rangle = \int_M h \, d\mu
\]
for $\mu$-almost every $z$. However, the convergence to this limit is at best as $1/T$ \cite{Kachurovskii96}
and can be arbitrarily slow \cite{Krengel78}.

To compute the average efficiently and accurately for a length-$T$ segment of an orbit,
we modify \Eq{Birkhoff} using the $C^\infty$ weight function\footnote
{Optimal choices for the weighting function have recently been explored by Ruth and Bindel using reduced rank extrapolation \cite{Ruth24}.}
\[
	\Psi(s) \equiv \left\{ \begin{array}{ll}  e^{-[s(1-s)]^{-1}}  & s \in (0,1) \\
	         								0	& s \le 0 \mbox{ or } s \ge 1 
						\end{array} \right. \;.
\]
This exponential bump function converges to zero with infinite smoothness 
at $0$ and $1$, i.e., $\Psi^{(k)}(0) = \Psi^{(k)}(1) = 0$ for all $k \in \bN$.
The finite-time weighted Birkhoff average (WBA) is then defined by
\beq{WB}
	\WB_{T}(h)(z) = \frac{1}{S}\sum_{t = 0}^{T-1} \Psi \left(\tfrac{t}{T}\right) h \circ f^t(z) \;, 
\eeq
with the normalization constant
\beq{SmoothedAve}
	S \equiv \sum_{t=0}^{T-1} \Psi \left(\tfrac{t}{T}\right)  \;. 
\eeq

As shown in \cite{Das16a}, \Eq{WB} gives the same answer as $T \to \infty$
as \Eq{Birkhoff}; however, for regular orbits it can converge much more quickly.
In particular, if the orbit is conjugate to a rigid rotation with a Diophantine rotation vector $\omega$
and the map $f$ and function $h$ are $C^\infty$, then \Eq{WB} converges faster than any power \cite{Das17}:
\[
	|\WB_T(h) - \langle h \rangle|  < \frac{c_k}{T^k}, \quad \forall k \in \bN .
\]

We estimate the error of the WBA for a given function $h$ and a given time $T$ by
computing the effective number of digits of accuracy:
\beq{digits}
	\digT = -\log_{10} \left|\WB_{T}(h)(z)-\WB_{T}(h)(f^T(z)) \right| ,
\eeq
i.e., comparing the result for the first $T$ iterates with that for the next $T$ iterates. In \cite{Sander20},
we observed that for the 2D Chirikov standard map, $\digT$ converges rapidly to machine precision for
orbits that lie on invariant circles: double precision accuracy ($\digT \sim 14$)
for most orbits is obtained within $T \approx 10^4$ iterates.
However, it converges slowly, or not at all for chaotic orbits. 
When the map has mixtures of regular and chaotic orbits, a histogram of $\mathrm{dig}_{10^4}$,
\cite[Fig. 3]{Sander20} shows two peaks, a
broader one around $\digT = 2$, corresponding to chaotic orbits and a
narrower one around $\digT = 14$, that corresponds to regular orbits. 
The latter peak has a tail over the interval $6 < \digT < 13$: for these orbits
the average converges more slowly. Such orbits typically 
lie near the boundary of narrow chaotic layers or are periodic with
high period within lower-period island chains. 
In \cite{Sander20} we showed our method is as accurate and more efficient than two standard methods for
identifying chaos: a positive Lyapunov exponent and the 0-1 test for chaos. 
In~\cite{Meiss21} we used the WBA to study two-tori in a 3D volume-preserving map. 

To obtain a criterion distinguishing chaotic and regular orbits, 
we choose a cutoff value for $\digT$, declaring that
\beq{ChaosThreshold}
	\digT < D_T \Rightarrow \mbox{ ``chaotic'' } ;
\eeq
conversely, all orbits with $\digT > D_T$ are ``nonchaotic''.
%
In \cite{Meiss21} we chose the cutoff $D_{10^6} = 11$
for a 3D map where the tori can be 1D or 2D. 
This cutoff is conservative in the sense that 
a chaotic orbit will be not be identified as regular, but there is a possibility of 
a regular orbit will be misidentified as chaotic. This choice had the  
benefit that the rotation vector of any orbit identified as regular can be computed with high accuracy,
and in that paper we were interested in studying the number theoretic properties of the robust tori.
For the current paper we use the less strict criterion
\beq{2DCriterion}
	T = 10^6, \quad D_T = 9 ,
\eeq
which is still  conservative in that chaotic orbits are quite unlikely to be identified as regular. 
 
\section{Resonance Orders}\label{app:ResonanceOrder}

In addition to providing the distinction between regular and chaotic orbits,
the  WBA can be used to compute an accurate value of the time average of a function $h$ of interest.
In particular, we can compute the rotation vector \Eq{RotationVector} of an orbit for a torus 
map of the form \Eq{torusMap} using \Eq{omegaWBA}.
If $T$ is large enough and the rotation vector exists, we expect
$\omega_T \approx \omega(x,f)$. Of course, for the quasiperiodic case,
\Eq{Omega2} implies that we only need to compute the first component of \Eq{omegaWBA}.

In general the resulting frequency vector can be rational, commensurate, or incommensurate.
To estimate whether a computed vector is essentially rational we would
ask that $\left|\omega-\tfrac{p}{q}\right|$ be small
for $p \in \bZ^d$, $q \in \bN$;
this would correspond to finding a rational approximation of the vector $\omega$.
More generally, $\omega \in \bR^d$ has a \textit{commensurability} 
(or is \textit{resonant} or \textit{mode-locked}) if
there is an $m \in \bZ^d \setminus\{0\}$ and an $n \in \bZ$ such that
\beq{ResonantPlanes}
	\omega \in \cR_{m,n} = \left\{\alpha \in \bR^d :  m \cdot \alpha =n \right\} ,
\eeq
i.e., it lies in the codimension-one plane $\cR_{m,n}$.  
We say that $\omega$ has \textit{resonance order} $M$ if it satisfies \Eq{ResonantPlanes}
and $M = \|m\|_1$ is the smallest length of such a (nonzero) vector $m$.
The set of vectors that do not lie in any plane are \textit{incommensurate};
an example is $\omega = (\sqrt{2},\sqrt{5})$.

For $d=2$, the sets $\cR_{m,n}$ are lines.
The \textit{rank} of the resonance for a given $\omega$ is the number of independent commensurability vectors $m$;
i.e., the dimension of the module of resonance vectors.
Note that $\omega$ is rational only if the rank is $d$; commensurabilities that have lower rank are partially
resonant, such as the rank-one vector $\omega = (3\sqrt{2},2\sqrt{2}-1)$ which lies in $\cR_{(2,-3),3}$ so that $M = 5$.

A vector $\omega$ is then \textit{approximately} commensurate if $\left| m\cdot \omega -n \right|$ is 
small, and in \cite{Meiss21} we developed a method for computing such commensurabilities. 
We say that a vector $\omega$ is $(m,n)$-resonant \textit{to precision $\delta$} if
the resonant plane intersects a ball of radius $\delta$ about $\omega$:
\beq{ResonancePlane}
	\cR_{m,n} \cap B_\delta(\omega) \neq \emptyset .
\eeq
Using the Euclidean norm, the minimum distance between the plane \Eq{ResonantPlanes}
and the point $\omega$ is
\beq{minDist}
	\Delta_{m,n}(\omega) = \min_{\alpha \in \cR_{m.n}} \| \alpha - \omega\|_2 
	                     = \frac{|m \cdot \omega-n|}{\|m\|_2} .
\eeq
Thus $\omega$ is $(m,n)$ resonant to precision $\delta$, whenever $\Delta_{m,n}(\omega) < \delta$, and
we call the value
\beq{Mres}
	M(\omega,\delta) = \min \{\|m\|_1 : \Delta_{m,n}(\omega) < \delta, \, m \in \bZ^d\setminus\{0\}, \, n\in \bZ \},
\eeq
the \textit{resonance order} of $\omega$.

As far as we know, there is no generalization of the $d=1$ Farey tree result 
used in \cite{Sander20} to compute \Eq{Mres} efficiently.\footnote
{The Kim-Ostlund tree  can be used to get resonance relations \cite{Ashwin93}; however, it is not clear that this algorithm returns a minimal $\|m\|$.} 
Nevertheless, since there are finitely many $m \in \bZ^d$ such that $\|m\|_1 \le M$, a brute force computation is of course possible for modest values of $M$; we gave such an algorithm in \cite{Meiss21}.

To understand what resonance orders are ``typical,'' in \cite{Meiss21} we computed the minimal resonance order
\Eq{Mres} for a set of equi-distributed, random $\omega \in [0,1]^2$ as a function of the precision $\delta$.
The resulting distribution of $\log(M)$ seen in \cite[Fig. 8]{Meiss21} has a mean and standard deviation
\bsplit{MeanRes}
	\langle \log_{10} M(\omega,\delta) \rangle &= -0.334\log_{10}(\delta) -0.091, \\
	\sigma &= 0.171.
\esplit
We observed that the standard deviation seems to be essentially independent of $\delta$, and 
we found a similar result for $d=1$ in \cite{Sander20}. Our computations inspired 
Chen and Haynes~\cite{Chen23} and more recently Marklof~\cite{marklof_log_2024, marklof_smallest_2024}
to find rigorous results for the resonance order distributions. 

Since the cutoff \Eq{2DCriterion} gives rotation number calculations 
accurate to within $10^{-9}$, we choose $\delta = 10^{-9}$ for \Eq{Mres}. For this case, \Eq{MeanRes} implies 
that $\langle\log_{10} M\rangle =  2.915$. We declare that a vector is nonresonant if 
\beq{Incommensurate}
	256 \le M \le 2673 \Rightarrow \mbox{``nonresonant''},
\eeq
corresponding to $ 2.407 < \log_{10}(M) < 3.427 $,
which is a range of approximately $\pm 3 \sigma$ about the mean \Eq{MeanRes}.
To test this criterion we selected $10^4$ uniformly randomly distributed values 
in $[0,1]^2$, and found that $1.36\%$ were incorrectly identified as resonant. 
Note that the distribution of log-orders for random vectors is not symmetric around the mean; 
in particular, $M<256$ occurred $1.32\%$ of the time, and $M>2673$ occurred $0.04\%$ of the time. 

We can further categorize the \textit{resonant} orbits (those 
that fail Criterion \Eq{Incommensurate}) by the rank of the resonance. 
Rank-two resonant orbits have frequencies on the intersection of a pair of different resonance lines.
For the quasiperiodic case we study here, $\omega_2 = \Omega_2$ is irrational;
therefore, all resonant orbits will have rank one.

\bibliographystyle{elsarticle-harv} 
\bibliography{TorusMaps}

\end{document}